\documentclass{lmcs}
\pdfoutput=1

\usepackage{lastpage}
\lmcsdoi{19}{4}{37}
\lmcsheading{}{\pageref{LastPage}}{}{}%
{Apr.~01,~2022}{Dec.~21,~2023}{}

\usepackage[utf8]{inputenc}
\usepackage{amsmath,amsfonts,amssymb,mathtools}
\usepackage{stmaryrd}
\usepackage{booktabs}
\usepackage{graphicx,float}
\usepackage[ruled,vlined]{algorithm2e}
\usepackage{algorithmic}
\usepackage{helvet}
\usepackage{courier}
\usepackage{multicol}
\usepackage{amssymb}
\usepackage{rotating}
\usepackage{epsfig}
\usepackage{latexsym}
\usepackage{graphicx}
\usepackage{color}
\usepackage{url}
\usepackage{amsfonts}
\usepackage{pgf}
\usepackage{tikz}
\usetikzlibrary{arrows,decorations.pathmorphing,backgrounds,positioning,fit,petri}
\usepackage{stmaryrd}
\usepackage{amsmath}
\usepackage{bussproofs}
\usepackage{amsthm}
\usepackage{xcolor}
\usepackage{centernot}

\usepackage[hidelinks]{hyperref}
\hypersetup{
    allcolors=black
}

\usepackage[capitalise, noabbrev]{cleveref}

\newcommand{\post}[1]{\mbox{$#1^{\bullet}$}}
\newcommand{\pre}[1]{\mbox{$^{\bullet}#1$}}

\newcommand{\deriv}[1]{{\mbox{${\:\stackrel{#1}{\longrightarrow}\:}$}}}
\newcommand{\derivp}[1]{{\mbox{${\:\stackrel{#1}{\rightarrowtail}\:}$}}}

\newcommand{\nderiv}[1]{\nrightarrow}

\renewcommand{\mid}{\;\;\big|\;\;}

\newcommand{\nat}{{\mathbb N}}

\newcommand{\trns}[1]{[{#1}\rangle}

\newcommand{\cnt}[0]{(\rho_{1}, C, \rho_{2})}
\newcommand{\cntp}[0]{(\rho_{1}', C', \rho_{2}')}
\newcommand{\fct}[0]{(\pi_1, f, \pi_2)}
\newcommand{\fctp}[0]{(\pi_1', f', \pi_2')}

\newcommand{\size}[1]{| \, #1 \, |}
\newcommand{\allim}[0]{\mathfrak{K}(S)}
\newcommand{\imtrans}[1]{\llbracket #1 \rangle}
\newcommand{\oim}[1]{(k_{#1}, \leq_{#1})}
\newcommand{\oimp}[1]{(k_{#1}', \leq_{#1}')}
\newcommand{\oims}[1]{(k_{#1}'', \leq_{#1}'')}

\newcommand{\kboilerplate}[1]{
    k'' \in k \boxminus \pre{#1} \text{ such that } 
    k' = k'' \boxplus \post{#1}
}
\newcommand{\kboilerplateindex}[2]{
    k_{#2}'' \in k_{#2} \boxminus \pre{#1} \text{ such that } 
    k_{#2}' = k_{#2}'' \boxplus \post{#1}
}
\newcommand{\oimct}[0]{(\oim{1}, \oim{2},\beta)}
\newcommand{\oimctp}[0]{(\oimp{1}, \oimp{2},\beta')}
\newcommand{\oimt}[0]{(\oim{1}, \oim{2},\beta)}
\newcommand{\oimtp}[0]{(\oimp{1}, \oimp{2},\beta')}
\newcommand{\oimmvs}[2]{\oim{#1}\imtrans{#2}\oimp{#1}}

\newcommand{\proctransp}[2]{(C, \rho_{#1}) \derivp{#2} (C', \rho_{#1}')}
\newcommand{\generated}[1]{k_{#1}' \setminus k_{#1}''}
\newcommand{\deleted}[1]{k_{#1} \setminus k_{#1}''}
\newcommand{\old}[1]{k_{#1}''}

\keywords{Behavioral equivalences, True concurrency, Fully-concurrent bisimilarity, Causal-net bisimilarity, i-Causal-net bisimilarity, Decidability}

\begin{document}

\title[Decidability of Two Truly Concurrent Equivalences]{Decidability of Two Truly Concurrent Equivalences \texorpdfstring{\\}{} for Finite Bounded Petri Nets}

\author[A. Cesco]{Arnaldo Cesco\lmcsorcid{0000-0002-3417-1890}}[a]
\author[R. Gorrieri]{Roberto Gorrieri\lmcsorcid{0000-0001-5502-0584}}[b]

\address{SECO Mind, Via Achille Grandi 20, 52100 Arezzo, Italy }
\email{arnaldo.cesco@secomind.com}

\address{Dipartimento di Informatica --- Scienza e Ingegneria
Universit\`a di Bologna, Mura A. Zamboni 7,
40127 Bologna, Italy}
\email{roberto.gorrieri@unibo.it}

\maketitle

\begin{abstract}
We prove that the well-known (strong) fully-concurrent bisimilarity and the novel i-causal-net bisimilarity, 
which is a sligtlhy coarser variant
of causal-net bisimilarity, are 
decidable for finite bounded Petri nets. 
The proofs are based on a generalization of
the ordered marking proof technique that Vogler used to demonstrate that (strong) fully-concurrent bisimilarity 
(or, equivalently, history-preserving bisimilarity)
is decidable on finite safe nets.
\end{abstract}

%
\section{Introduction} 
%

The causal semantics of Petri nets has been studied according to the 
so-called {\em individual token interpretation}, where multiple tokens on the same place are seen as individual entities,
starting from the work of Goltz and Reisig \cite{GR83}, further explored and extended in \cite{Eng91,MMS97}.
However, the token game on such nets is defined according to the so-called {\em collective token interpretation}, where
multiple tokens on the same place are considered as indistinguishable. Causal semantics for Petri nets under this more relaxed
interpretation is under investigation (see, e.g., \cite{BD87,MM90} for important contributions in the linear-time semantics), but a 
completely satisfactory solution to the general problem is not yet available (see the survey \cite{vGGS11}
and the references therein).

The main aim of this paper is to investigate the decidability of two behavioral equivalences defined over the
causal semantics developed for Petri nets under the
individual token interpretation, based on the concept of {\em causal net} \cite{GR83,Eng91,MMS97}. 
In particular, we study the well-known {\em fully-concurrent bisimilarity} \cite{BDKP91} and a variant of
{\em causal-net bisimilarity} \cite{vG15,Gor22} (we call {\em i-causal-net bisimilarity}), which have been advocated
as very suitable  equivalences to compare the behavior of Petri nets.

Fully-concurrent bisimilarity considers as equivalent those markings that
can perform the same partial orders of events. Its definition
was inspired by previous notions of equivalence on other models of concurrency:
{\em history-preserving bisimulation}, originally defined in \cite{RT88} under the name of {\em behavior-structure bisimulation}, and 
then elaborated on in \cite{vGG89} (who called it by this name) and also independently defined in \cite{DDM89} 
(who called it by {\em mixed ordering bisimulation}).
Causal-net bisimilarity \cite{vG15,Gor22}, which coincides with {\em structure-preserving bisimilarity} \cite{vG15}, observes 
not only the partial order of events, but also the size of the distributed state: more precisely, if two markings are related by a causal-net bisimulation, 
then they generate the same causal nets.

We think that causal-net bisimilarity is more accurate than 
fully-concurrent bisimilarity because it   
is {\em resource-aware}. In fact, in the implementation of a system, a token is an instance of a 
sequential process, so that a processor is needed to execute it. If two markings 
have different size, then a different number of processors is necessary. 
Hence, a semantics such as causal-net bisimilarity, which relates markings of the same size only, is more accurate
as it equates distributed systems only if they require the same amount of execution resources.
Van Glabbeek \cite{vG15} argued that structure-preserving bisimilarity (hence, also causal-net bisimilarity) 
is the most appropriate behavioral equivalence for Petri nets, as it is the only one
respecting a list of 9 desirable requirements he proposed. 
Moreover, \cite{Gor20} offers, in the area of information flow security, further arguments in favor of concrete, resource-aware
equivalences that, differently
from fully-concurrent bisimilarity, observe
also the size of the distributed state. 

However, we think that causal-net bisimilarity is a bit too discriminating, because of a peculiar aspect in its definition: 
the bijective mappings from the current marking of the causal net to the two markings under scrutiny
are defined {\em before} the bisimulation game starts,
and this may impede to equate some markings generating the same causal 
nets. (An illustrative example explaining this shortcoming 
is outlined in Example \ref{icn<cn-ex}.)
For this reason, we propose in this paper
a variant definition, where the bijective mappings of the current markings are not defined (i.e., the 
mappings of the maximal conditions of the causal net are left open)
{\em before} the bisimulation game starts; on the contrary, they are
actually constructed progressively as long as the computation proceeds. More precisely, the selection of the matching transitions is 
made {\em before} fixing the mapping, or equivalently, the mapping is partially computed {\em after} choosing the matching transitions.
This variant bisimilarity, we call {\em i-causal-net bisimilarity} (where the prefix i- denotes that 
it works up to the choice of the $i$somorphism of the current markings), is slightly coarser than causal-net bisimilarity,
but still ensuring that related markings generate the same causal nets, so that it is resource-aware, too,
even if it does not respect all the desiderata listed by van Glabbeek in \cite{vG15} (e.g., it does not respect {\em inevitability} \cite{MOP89}, i.e., 
if two systems are equivalent, and in one the occurrence of a certain action is inevitable, then so is it in the other one.)
The definition of i-causal-net bisimilarity is the first contribution of this paper.

The results about the decidability of truly concurrent behavioral equivalences are rather rare \cite{Vog91,JM96,Vog95,MP97}
and are often limited to the class of finite safe nets, i.e., nets whose places can hold one token 
at most. Our main aim is to extend some of these decidability proofs to the case of bounded nets.

In his seminal paper \cite{Vog91},
Vogler demonstrated that (strong) fully-concurrent bisimilarity is decidable on finite safe nets. His proof is based 
on an alternative characterization of fully-concurrent
bisimulation, called {\em ordered marking bisimulation} (OM bisimulation, for short), which is based on the idea of 
representing the current global state of the net system as a marking 
equipped with a pre-ordering on its tokens, that reflects the causal ordering of the transitions that produced the tokens. 
However, the ordered marking idea works well if the marking is a set (as it is the case for safe nets), and so it is not immediate to generalize it to bounded
 nets, whose markings are, in general, multisets.
 
 The second contribution of this paper is the definition of an alternative token game semantics for Petri nets which is 
 defined according to the {\em individual token
 philosophy}, rather than the {\em collective token philosophy}, as it is customary for Petri nets. 
Differently from the first, rather complex, definition of this sort, proposed in \cite{vG05}, 
 we achieve this goal by 
 representing each token simply as a pair $(s, i)$, 
 where $s$ is the name of the place where the token is on, and $i$ is a natural number (an index) 
 assigned to the token in such a way that different tokens 
 on the same place have different indexes. In this way, a multiset over the set of places (i.e., a marking) is turned into a set of indexed places.
 The main advantage of having turned multisets into sets is that  Vogler's ordered marking idea can be used
 also in this richer context, yielding {\em  ordered indexed markings}.

The third contribution of the paper is to show that (strong) fully-concurrent bisimulation can be equivalently characterized
as a suitable bisimulation over ordered indexed markings, called {\em OIM bisimulation}, generalizing the approach by
Vogler \cite{Vog91}. An OIM bisimulation is formed by a set of triples, each 
composed of two ordered indexed markings
and a relation between these two  ordered indexed markings that respects the pre-orders.
The decidability of (strong) fully-concurrent
bisimilarity on finite bounded nets follows by observing that the reachable indexed markings are finitely many, so the ordered
indexed markings of interest are finitely many as well, so that there can only be finitely many candidate relations (which are all finite) to be OIM bisimulations.

The fourth contribution of the paper is to show that our generalization of Vogler's proof technique can be adapted to prove the
decidability on bounded finite nets also of {\em i-causal-net bisimilarity}. This result is obtained by showing that it
can be equivalently characterized
as a suitable bisimulation over ordered indexed markings, called {\em OIMC bisimulation}, which is a variant of OIM bisimulation
with similar finiteness properties. 

The paper is organized as follows. Section~\ref{def-sec} recalls the basic definitions about Petri nets. 
Section~\ref{causal-sem-sec}
recalls the causal semantics, including the definition of causal-net bisimilarity and (strong) fully-concurrent bisimilarity,
and proposes the novel i-causal-net bisimilarity.
Section~\ref{index-sec} introduces indexed markings and the alternative token game semantics according to the individual token philosophy. 
Section~\ref{oim-sec} describes indexed ordered markings and their properties.
Section~\ref{fc-dec-sec} introduces OIM bisimulation, proves that its equivalence coincides with (strong) fully-concurrent bisimilarity and, moreover,
shows that it is decidable.
Section~\ref{icn-dec-sec} proves that also i-causal-net bisimilarity is decidable.
Finally, in Section~\ref{conc-sec} we discuss related literature, we hint that other 
truly concurrent behavioral equivalences are decidable for finite bounded Petri nets and we also suggest some future research.\\

{\em A preliminary version of this paper appeared in the proceedings of the 22nd Italian Conference on Theoretical Computer Science, Bologna, Italy, September 13-15, 2021 \cite{CG21}.}

%
\section{Basic Definitions} \label{def-sec}
%

\begin{defi}\label{multiset}{\bf (Multiset)}
Let $\nat$ be the set of natural numbers. 
Given a finite set $S$, a {\em multiset} over $S$ is a function $m: S \rightarrow\nat$. 
The {\em support} set $dom(m)$ of $m$ is $\{ s \in S \mid m(s) \neq 0\}$. 
The set of all multisets 
over $S$,  denoted by ${\mathcal M}(S)$, is ranged over by $m$. 
We write $s \in m$ if $m(s)>0$. 
The {\em multiplicity} of $s$ in $m$ is given by the number $m(s)$. The {\em size} of $m$, denoted by $|m|$,
is the number $\sum_{s\in S} m(s)$, i.e., the total number of its elements.
A multiset $m$ such 
that $dom(m) = \emptyset$ is called {\em empty} and is denoted by $\theta$.
We write $m \subseteq m'$ if $m(s) \leq m'(s)$ for all $s \in S$. 
{\em Multiset union} $\_ \oplus \_$ is defined as: $(m \oplus m')(s)$ $ = m(s) + m'(s)$; it is commutative, associative and has $\theta$ as neutral element. 
{\em Multiset difference} $\_ \ominus \_$ is defined as: 
$(m_1 \ominus m_2)(s) = max\{m_1(s) - m_2(s), 0\}$.
The {\em scalar product} of a number $j$ with $m$ is the multiset $j \cdot m$ defined as
$(j \cdot m)(s) = j \cdot (m(s))$. By $s_i$ we also denote the multiset with $s_i$ as its only element.
Hence, a multiset $m$ over $S = \{s_1, \ldots, s_n\}$
can be represented as $k_1\cdot s_{1} \oplus k_2 \cdot s_{2} \oplus \ldots \oplus k_n \cdot s_{n}$,
where $k_j = m(s_{j}) \geq 0$ for $j= 1, \ldots, n$.
\end{defi}

\begin{defi}\label{pt-net-def}{\bf (Place/Transition net)}
A labeled {\em Place/Transition} Petri net (P/T net for short) is a tuple $N = (S, A, T)$, where
\begin{itemize}
\item 
$S$ is the finite set of {\em places}, ranged over by $s$ (possibly indexed),
\item 
$A$ is the finite set of {\em labels}, ranged over by $\ell$ (possibly indexed), and
\item 
$T \subseteq {(\mathcal M}(S) \setminus \{\theta\}) \times A \times {\mathcal M}(S)$ 
is the finite set of {\em transitions}, 
ranged over by $t$ (possibly indexed).
\end{itemize}
Given a transition $t = (m, \ell, m')$,
we use the notation:
\begin{itemize}
\item  $\pre t$ to denote its {\em pre-set} $m$ (which cannot be an empty multiset) of tokens to be consumed;
\item $l(t)$ for its {\em label} $\ell$, and
\item $\post t$ to denote its {\em post-set} $m'$ of tokens to be produced.
\end{itemize}
Hence, transition $t$ can be also represented as $\pre t \deriv{l(t)} \post t$.
We also define the {\em flow function}
$\mathsf{flow}: (S \times T) \cup (T \times S) \rightarrow \nat$ as follows:
for all $s \in S$, for all $t \in T$,
$\mathsf{flow}(s,t) = \pre{t}(s)$ and $\mathsf{flow}(t,s) = \post{t}(s)$
(note that $\pre{t}(s)$ and $\post{t}(s)$ are integers, representing the multiplicity of $s$ in $\pre{t}$ and $\post{t}$, respectively).
We will use $F$ to denote the {\em flow relation} 
$\{(x,y) \mid x,y \in S \cup T \, \wedge \, \mathsf{flow}(x,y) > 0\}$.
Finally, we define pre-sets and post-sets also for places as follows: $\pre s = \{t \in T \mid s \in \post t\}$
and $\post s = \{t \in T \mid s \in \pre t\}$. 
\end{defi}

In the graphical description of finite P/T nets, 
places (represented as circles) and
transitions (represented as boxes) 
are connected by directed arcs.
The arcs may be labeled with the number representing how many tokens
of that type are to be removed from (or added to) that place, as specified by function $\mathsf{flow}$; no label on
the arc is interpreted as the number one, i.e., one token flowing on the arc. 
This numerical label of the arc is called its {\em weight}.

\begin{defi}\label{net-system}{\bf (Marking, P/T net system)}
A multiset over $S$  is called a {\em marking}. Given a marking $m$ and a place $s$, 
we say that the place $s$ contains $m(s)$ {\em tokens}, graphically represented by $m(s)$ bullets
inside place $s$.
A {\em P/T net system} $N(m_0)$ is a tuple $(S, A, T, m_{0})$, where $(S,A, T)$ is a P/T net and $m_{0}$ is  
a marking over $S$, called
the {\em initial marking}. We also say that $N(m_0)$ is a {\em marked} net.
\end{defi}

The sequential semantics of a marked net is defined by the so-called {\em token game}, 
describing the flow of tokens through it. 
There are several possible variants of the token game (see, e.g., \cite{vG05}). 
Below we present the standard token game, following the so-called {\em collective interpretation}, according to which multiple tokens on the 
same place are indistinguishable, while in Section~\ref{index-sec} we introduce a novel variant following the
so-called {\em individual interpretation}.

\begin{defi}\label{token-game} {\bf (Token game)}
A transition $t $ is {\em enabled} at $m$, denoted $m[t\rangle$, if $\pre t \subseteq m$. 
The {\em firing} of  $t$ enabled at $m$ produces the marking 
$m' = (m \ominus  \pre t) \oplus \post t$, written $m[t\rangle m'$. 
\end{defi}

\begin{defi}\label{firing-reachable}{\bf (Firing sequence, reachable marking)}
A {\em firing sequence} starting at $m$ is defined inductively as follows:
\begin{itemize}
\item $m[\epsilon\rangle m$ is a firing sequence (where $\epsilon$ denotes an empty sequence of transitions) and
\item if $m[\sigma\rangle m'$ is a firing sequence and $m' [t\rangle m''$, then
$m [\sigma t\rangle m''$ is a firing sequence. 
\end{itemize}
If $\sigma = t_1 \ldots t_n$ (for $n \geq 0$) and $m[\sigma\rangle m'$ is a firing sequence, then there exist $m_1,  \ldots, m_{n+1}$ such that
$m = m_1[t_1\rangle m_2 [t_2\rangle \ldots 
\ldots m_n [t_n\rangle m_{n+1} = m'$, and 
$\sigma = t_1 \ldots t_n$ is called a {\em transition sequence} starting at $m$ and ending 
at $m'$. The set of {\em reachable markings} from $m$ is 
$[m\rangle = \{m' \mid \exists \sigma.
m[\sigma\rangle m'\}$. 
Note that the set of reachable markings may be countably infinite for finite P/T nets.
\end{defi}

\begin{defi}\label{net-classes}{\bf (Classes of finite P/T Nets)}
A {\em finite marked}  P/T net $N = (S, A, T, m_0)$ is: 
\begin{itemize}
    \item {\em safe} if each place contains at most one token in each reachable marking, i.e.,  
    $\forall m \in [m_0 \rangle,\forall s \in S,$ $ m(s) \leq 1 $.
   
    \item {\em bounded} if the number of tokens in each place is bounded by some $k$ for each reachable marking, i.e., $\exists k \in \nat$ such that 
    $ \forall m \in [m_0 \rangle , \forall s \in S$ we have that $m(s) \leq k$.
    If this is the case, we say that the net is $k$-bounded (hence, a safe net is just a 1-bounded net). 
\end{itemize} 
\end{defi}

%
\section{Causality-based Semantics}\label{causal-sem-sec}
%

We first outline some definitions (in particular, {\em causal net}, {\em folding}  and {\em process}), adapted from the literature
(cf., e.g., \cite{GR83,BD87,Old91,Vog91,vG15,Gor22}), that are necessary to introduce causal-net bisimilarity \cite{vG15,Gor22}
and fully-concurrent bisimilarity \cite{BDKP91}.
Then we propose some novel definitions (in particular, {\em partial folding}
and {\em partial process}) that are necessary to introduce the novel i-causal-net bisimilarity.

%
\subsection{Causal Nets and Processes}\label{cn-proc-sec}
%

\begin{defi}\label{acyc-def}{\bf (Acyclic net)}
A P/T net $N = (S, A, T)$ is
 {\em acyclic} if its flow relation $F$ is acyclic (i.e., $\not \exists x$ such that $x F^+ x$, 
 where $F^+$ is the transitive closure of $F$).
\end{defi}

The concurrent semantics of a marked P/T net is defined by a particular class of  acyclic safe nets, 
where places are not branched (hence they represent a single run) and all arcs have weight 1. 
This kind of net is called {\em causal net}. 
We use the name $C$ (possibly indexed) to denote a causal net, the set $B$ to denote its 
places (called {\em conditions}), the set $E$ to denote its transitions 
(called {\em events}), and
$L$ to denote its labels.

\begin{defi}\label{causalnet-def}{\bf (Causal net)}
A causal net is a finite marked net $C(\mathsf{m}_0) = (B,L, 
E,  \mathsf{m}_0)$ satisfying
the following conditions:
\begin{enumerate}
\item $C$ is acyclic;
\item $\forall b \in B \; \; | \pre{b} | \leq 1\, \wedge \, | \post{b} | \leq 1$ (i.e., the places are not branched);
\item  $ \forall b \in B \; \; \mathsf{m}_0(b)   =  \begin{cases}
 1 & \mbox{if $\; \pre{b} = \emptyset$}\\ 
  0  & \mbox{otherwise;}   
   \end{cases}$\\
\item $\forall e \in E \; \; \pre{e}(b) \leq 1 \, \wedge \, \post{e}(b) \leq 1$ for all $b \in B$ (i.e., all the arcs have weight $1$).
\end{enumerate}
We denote by $Min(C)$ the set $\mathsf{m}_0$, and by $Max(C)$ the set
$\{b \in B \mid \post{b} = \emptyset\}$.
\end{defi}

Note that any reachable marking of a causal net is a set, i.e., 
this net is {\em safe}; in fact, the initial marking is a set and, 
assuming by induction that a reachable marking $\mathsf{m}$ is a set and enables $e$, i.e., 
$\mathsf{m}[e\rangle \mathsf{m}'$,
then also
$\mathsf{m}' =  (\mathsf{m} \ominus \pre{e}) \oplus \post{e}$ is a set, 
as the net is acyclic (with unbranched places) and because
of the condition on the shape of the post-set of $e$ (weights can only be $1$).

 As the initial marking of a causal net is fixed by its shape (according to item $3$ of 
 Definition \ref{causalnet-def}), in the following, in order to make the 
 notation lighter, we often omit the indication of the initial marking (also in their graphical representation), 
 so that the causal 
 net $C(\mathsf{m}_0)$ is simply denoted by $C$.

\begin{defi}\label{trans-causal}{\bf (Moves of a causal net)}
Given two causal nets $C = (B, L, E,  \mathsf{m}_0)$
and $C' = (B', L,$ $ E',  \mathsf{m}_0)$, we say that $C$
moves in one step to $C'$ through $e$, denoted by
$C [e\rangle C'$, if $\; \pre{e} \subseteq Max(C)$, $E' = E \cup \{e\}$
and $B' = B \cup \post{e}$; in other words,  $C'$ extends $C$ by one event $e$.
\end{defi}

\begin{defi}\label{folding-def}{\bf (Folding and Process)}
A {\em folding} from a causal net $C = (B, L, E, \mathsf{m}_0)$ into a net system
$N(m_0) = (S, A, T, m_0)$ is a function $\rho: B \cup E \to S \cup T$, which is type-preserving, i.e., such that $\rho(B) \subseteq S$ and $\rho(E) \subseteq T$, satisfying the following:
\begin{itemize}
\item $L = A$ and $\mathsf{l}(e) = l(\rho(e))$ for all $e \in E$;
\item $\rho(\mathsf{m}_0) = m_0$, i.e., $m_0(s) = | \rho^{-1}(s) \cap \mathsf{m}_0 |$;
\item $\forall e \in E, \rho(\pre{e}) = \pre{\rho(e)}$, i.e., $\rho(\pre{e})(s) = | \rho^{-1}(s) \cap \pre{e} |$
for all $s \in S$;
\item $\forall e \in E, \, \rho(\post{e}) = \post{\rho(e)}$,  i.e., $\rho(\post{e})(s) = | \rho^{-1}(s) \cap \post{e} |$
for all $s \in S$.
\end{itemize}
A pair $(C, \rho)$, where $C$ is a causal net and $\rho$ a folding from  
$C$ to a net system $N(m_0)$, is a {\em process} of $N(m_0)$, written also as $\pi$.
\end{defi}

\begin{defi}\label{po-process-def}{\bf (Partial orders of events from a process)}
From a causal net $C = (B, L, E, \mathsf{m}_0)$,  we can 
extract the {\em partial order of its events}
$\mathsf{E}_{\mathsf{C}} = (E, \preceq)$,
where $e_1 \preceq e_2$ if there is a path in the net from $e_1$ to $e_2$, i.e., if $e_1 \mathsf{F}^* e_2$, where
$\mathsf{F}^*$ is the reflexive and transitive closure of
$\mathsf{F}$, which is the flow relation for $C$.
\noindent
Given a process $\pi = (C, \rho)$, we denote $\preceq$ as $\leq_\pi$, 
i.e., given $e_1, e_2 \in E$, $e_1 \leq_\pi e_2$ if and only if $e_1 \preceq e_2$.
\noindent
Given two partial orders of events $\mathsf{E}_{\mathsf{C}_1} = (E_1, \preceq_1)$ and $\mathsf{E}_{\mathsf{C}_2} = (E_2, \preceq_2)$,
we say that they are isomorphic if there exists an order-preserving bijection $f: E_1 \rightarrow E_2$, i.e., such that 
$e_1 \preceq_1 e_2$ if and only if $f(e_1) \preceq_2 f(e_2)$. In such a case, we say that $f$ is an {\em isomorphism} between $\mathsf{E}_{\mathsf{C}_1}$
and $\mathsf{E}_{\mathsf{C}_2}$.
\end{defi}

\begin{defi}\label{trans-process}{\bf (Moves of a process, event sequence)}
Let $N(m_0) = (S, A, T, m_0)$ be a net system 
and let $(C_i, \rho_i)$, for $i = 1, 2$, be two processes of $N(m_0)$.
We say that $(C_1, \rho_1)$
moves in one step to $(C_2, \rho_2)$ through $e$, denoted by
$(C_1, \rho_1) \deriv{e} (C_2, \rho_2)$, if $C_1 [e\rangle C_2$
and $\rho_1 \subseteq \rho_2$.
If $\pi_1 = (C_1, \rho_1)$ and $\pi_2 = (C_2, \rho_2)$, we denote
the move as $\pi_1 \deriv{e} \pi_2$.

An {\em event sequence} starting at a process $\pi$ is defined as follows:
\begin{itemize}
\item $\pi[\epsilon\rangle \pi$ is an event sequence (where $\epsilon$ denotes an empty sequence of events) and
\item if $\pi[\sigma\rangle \pi'$ is an event sequence and $\pi' \deriv{e} \pi''$, then
$\pi [\sigma e\rangle \pi''$ is an event sequence. 
\end{itemize}
\end{defi}

\begin{prop} \label{preset-not-eq-pi}

Assume that $\pi = (C, \rho)$ is a process of $N(m_0)$ 
such that $\pi \deriv{e} \pi' = (C', \rho')$,
i.e. $\pi$ moves in one step trough $e$ to $\pi'$.
Then, 
$\forall b \in Max(C) \, , \,
\forall b' \in \post{e} , \,$
if  $\pre{b} \leq_{\pi'} \pre{b'}$, then $\exists b'' \in \pre{e}$
such that 
$\pre{b} \leq_\pi \pre{b''}$.

\proof
    By Definition \ref{po-process-def}, $\pre{b} \leq_{\pi'} \pre{b'}$ 
    means that there exists a path in $C'$ 
    starting from
    $\pre{b}$ and ending at $\pre{b'}$.
    Let us choose $b''$ to be the condition 
    immediately before $\pre{b'}$ in that path.
    It follows that there exists a path in $C$
    starting from $\pre{b}$ and ending at $\pre{b''}$:
    then, by Definition \ref{po-process-def},
    we get the thesis.
\qed
\end{prop}

%
\subsection{Causal-net Bisimilarity and Fully-concurrent Bisimilarity}\label{cn-fc-sec}
%

We now recall the definition of causal-net bisimulation \cite{vG15,Gor22},
a process-based equivalence relating both the history of two executions and
their (distributed) state. 

\begin{defi}\label{cn-bis-def}{\bf (Causal-net bisimulation)}
Let $N = (S, A, T)$ be a finite P/T net. A {\em causal-net bisimulation} 
is a relation $R$, composed of 
triples of the form $(\rho_1, C, \rho_2)$, where, for $i = 1, 2$, $(C, \rho_i)$ is a process of $N(m_{0i})$ for some $m_{0i}$,
such that if $(\rho_1, C, \rho_2) \in R$ then

\begin{itemize}[align=left]
\item[$i)$] 
$\forall t_1, C', \rho_1'$ such that $(C, \rho_1) \deriv{e}$ \\ $(C', \rho_1')$,
where $\rho_1'(e) = t_1$,
$\exists t_2, \rho_2'$ such that
$(C, \rho_2) \deriv{e} (C', \rho_2')$,
where $\rho_2'(e) = t_2$, and
$(\rho'_1, C', \rho'_2) \in R$;

\item[$ii)$] symmetrically, 
$\forall t_2, C', \rho_2'$ such that $(C, \rho_2) \deriv{e} (C', \rho_2')$,
where $\rho_2'(e) = t_2$,
$\exists t_1, \rho_1'$ such that \\
$(C, \rho_1) \deriv{e} (C', \rho_1')$,
where $\rho_1'(e) = t_1$, and
$(\rho'_1, C', \rho'_2) \in R$.

\end{itemize}

\noindent
Two markings $m_{1}$ and $m_2$ of $N$ are cn-bisimilar (or cn-bisimulation equivalent), 
denoted by $m_{1} \sim_{cn} m_{2}$, 
if there exists a causal-net bisimulation $R$ containing a triple $(\rho^0_1, C^0, \rho^0_2)$, 
where $C^0$ contains no events and 
$\rho^0_i(Min( C^0))  = \rho^0_i(Max( C^0)) = m_i\;$ for $i = 1, 2$.
\end{defi}

Causal-net bisimilarity \cite{vG15,Gor22}, which coincides with {\em structure-preserving bisimilarity} \cite{vG15}, observes 
not only the partial orders of events, but also the size of the distributed state; in fact, it observes the causal nets.
A weaker equivalence, observing only the partial orders of the events performed, is 
{\em fully-concurrent bisimulation} (fc-bisimulation, for short) \cite{BDKP91}.
Here we present the strong version, where all the events are considered observable.

\begin{defi} \label{fc-bis} {\bf (Fully-concurrent bisimilarity)}
Given a finite P/T net $N = (S, A, T)$, a {\em fully-con-current bisimulation}
is a relation $R$, composed of triples of the form $\fct$ where, 
for $i = 1,2$, 
$\pi_i = (C_i, \rho_i)$ is a process of $N(m_{0i})$ for some $m_{0i}$ and
$f$ is an isomorphism between $\mathsf{E}_{C_1}$ and $\mathsf{E}_{C_2}$,
such that if $\fct \in R$ then:
\begin{itemize}[align=left]
    \item[$i)$] $\forall t_1, e_1, \pi_1'$ such that $\pi_1 \deriv{e_1} \pi_1'$,
    where $\rho_1'(e_1) = t_1$, there exist $e_2, t_2, \pi_2', f'$ such that 
        \begin{enumerate}
            \item $\pi_2 \deriv{e_2} \pi_2'$ 
                where $\rho_2'(e_2) = t_2$ and $l(t_1) = l(t_2)$,
                \item $f' = f \mathbin{\mathaccent\cdot\cup} \{ e_1 \mapsto e_2 \}$,
            \item $\fctp \in R$;
        \end{enumerate}
    \item[$ii)$] symmetrically, if $\pi_2$ moves first.
\end{itemize}

\noindent
Two markings $m_1, m_2$ of $N$ are fc-bisimilar,
denoted by $m_1 \sim_{fc} m_2$ if a fully-concurrent bisimulation R exists,
containing a triple $(\pi^0_1, \emptyset, \pi^0_2)$ where 
$\pi^0_i = (C^0_i, \rho^0_i)$ is such that 
$C^0_i$ contains no events and 
$\rho^0_i(Min(C^0_i))  = \rho^0_i(Max(C^0_i))$ $ = m_i\;$ for $i = 1, 2$.
\end{defi}

Of course, $\sim_{cn}$ is finer than $\sim_{fc}$. This can be proved \cite{vG15} by observing that 
if $R_1$ is a causal-net bisimulation,
then $R_2 = \{(C, \rho_1), id, (C, \rho_2) \mid (\rho_1, C, \rho_2) \in R_1\}$, where $id$ is 
the identity function on $E$, is an fc-bisimulation.
This implication is strict, as illustrated by the following example.

\begin{figure}[t]
    \centering
    
    \begin{tikzpicture}[
        every place/.style={draw,thick,inner sep=0pt,minimum size=6mm},
        every transition/.style={draw,thick,inner sep=0pt,minimum size=4mm},
        bend angle=30,
        pre/.style={<-,shorten <=1pt,>=stealth,semithick},
        post/.style={->,shorten >=1pt,>=stealth,semithick}    
    ]
    
    \node (a) [label=left:$N)\qquad$]{};
    \node (p1) [place]  [label=above:$s_1$] {};   
    \node (t1) [transition] [below of = p1,label=left:$a$] {};
    \node (p2) [place] [below of = t1,label=left:$s_2$] {};
    \node (p4) [place]  [right = {1cm} of p1, label=above:$s_3$] {};
    \node (t4) [transition] [below of = p4,label=left:$a$] {};
    
    \draw  [->] (p1) to (t1);
    \draw  [->] (t1) to (p2);
    \draw  [->] (p4) to (t4);
    
    \node (b) [right = {2cm} of p4, label=left:$C_1)$]{};
    \node (b1) [place]  [right = {0.2cm} of b, label=above:$b_1$] {};   
    \node (e1) [transition] [below of = b1,label=left:$e_{a_1}$] {};
    \node (b2) [place] [below of = e1,label=left:$b_2$] {};
    
    \draw  [->] (b1) to (e1);
    \draw  [->] (e1) to (b2);

    \node (bp) [right = {2cm} of b1, label=left:$C_2)$]{};
    \node (b3) [place]  [right = {0.2cm} of bp, label=above:$b_3$] {};
    \node (e3) [transition] [below of = b3,label=left:$e_{a_2}$] {};
    
    \draw  [->] (b3) to (e3);
    
    \end{tikzpicture}
    
    \caption{A finite P/T net $N$ and two causal nets: $C_1$ corresponds to the maximal process of $N(s_1)$ and $C_2$ corresponds to the maximal process of $N(s_3)$.}
    \label{fig:cn_bis_fc_bis_example}
\end{figure}

\begin{exa}\label{cn_vs_fc_example}
In Figure \ref{fig:cn_bis_fc_bis_example} a simple finite P/T net $N$ is given.
It is easy to see that $C_1$ (resp. $C_2$) corresponds to a process $\pi_1$ (resp. $\pi_2$) of $N(s_1)$ (resp. $N(s_3)$), 
where $\rho_1$ (resp. $\rho_2$) maps each condition $b_i$ to the place $s_i$ having the same subscript and 
each event to the corresponding transition having the same shape.  
In the graphical depiction of causal nets, we will omit the initial marking 
for simplicity, since it can be inferred by looking at conditions of the causal net with empty preset.

\noindent
Consider places $s_1$ and $s_3$: we have $s_1 \sim_{fc} s_3$ and this is proved by relation
\begin{center}
    $R = \{ (
            ((b_1, \{ a \}, \emptyset, b_1), b_1 \mapsto s_1), 
            \emptyset, 
            ((b_3, \{ a \}, \emptyset, b_3), b_3 \mapsto s_3)), 
           (\pi_1, e_{a_1} \mapsto e_{a_2}, \pi_2) \}$.
\end{center}
Indeed, $((b_1, \{ a \}, \emptyset, b_1), b_1 \mapsto s_1)$ is a process of $N(s_1)$ and
$((b_3, \{ a \}, \emptyset, b_3), b_3 \mapsto s_3)$ is a process of $N(s_3)$, as
both processes contain no events and are such that minimal and maximal conditions are the same and mapped on the corresponding initial markings. 
If $((b_1, \{ a \}, \emptyset, b_1), b_1 \mapsto s_1)$ moves first by $((b_1, \{ a \}, \emptyset, b_1), b_1 \mapsto s_1) \deriv{e_{a_1}} \pi_1$, then the other process
$((b_3, \{ a \},$ $ \emptyset, b_3)$, $b_3 \mapsto s_3)$ can respond with $((b_3, \{ a \}, \emptyset, b_3), b_3 \mapsto s_3) \deriv{e_{a_2}} \pi_2$, and
$(\pi_1, e_{a_1} \mapsto e_{a_2},$ $ \pi_2) \in R$. 
Symmetrical is the case when $((b_3, \{ a \}, \emptyset, b_3), b_3 \mapsto s_3)$ moves first, and so it is omitted.

However, it is not true that $s_1 \sim_{cn} s_3$, because 
$C_1$ and $C_2$ are not isomorphic and therefore it is not possible to build a causal-net bisimulation.
\end{exa}

%
\subsection{I-causal-net Bisimilarity}\label{icn-bis-sec}
%

Causal-net bisimulation may be criticized because it may fail to equate nets that, intuitively, should be considered 
equivalent, as they can perform the same
causal nets, as illustrated in the following example.

\begin{exa}\label{icn<cn-ex}
Consider the nets in Figure \ref{cn-vs-icn-fig} and the two markings $s_1 \oplus s_2 \oplus s_3$ and $r_1 \oplus r_2 \oplus r_3$. 
Let us consider the initial causal net $C^0$ composed of three conditions $b_1, b_2, b_3$ only.
Whatever are the initial mappings $\rho_1^0$ and $\rho_2^0$ from conditions to places, it is always possible for the first net to perform a transition that is not matched by the second net.
For instance, assume that these mappings are the trivial ones, i.e., $\rho_1^0$ maps $b_i$ to $s_i$ and $\rho_2^0$ maps $b_i$ to $r_i$ for $i = 1, 2, 3$. Then, if the first net performs
the transition $(s_1 \oplus s_3, a, \theta)$, the second net cannot reply because $r_1 \oplus r_3$ is stuck. However, these two nets should be considered equivalent, because
what they can do is just one single causal net, which is isomorphic to the one on the right of Figure \ref{cn-vs-icn-fig}.
\end{exa}

\begin{figure}[t]
\centering

\begin{tikzpicture}[
every place/.style={draw,thick,inner sep=0pt,minimum size=6mm},
every transition/.style={draw,thick,inner sep=0pt,minimum size=4mm},
bend angle=42,
pre/.style={<-,shorten <=1pt,>=stealth,semithick},
post/.style={->,shorten >=1pt,>=stealth,semithick}
]
\def\eofigdist{3cm}
\def\eodist{0.5cm}
\def\eodisty{1.5cm}

\node (p1) [place,tokens=1]  [label=above:$s_1$] {};
\node (p2) [place,tokens=1]  [right=\eodist of p1,label=above:$s_2$] {};
\node (p3) [place,tokens=1] [right=\eodist of p2,label=above:$s_3$] {};
\node (t1) [transition] [below=\eodist of p1,label=left:$a$] {};
\node (t2) [transition] [below=\eodist of p2,label=left:$a$] {};
\node (t3) [transition] [below=\eodist of p3,label=right:$a$] {};

\draw  [->] (p1) to (t1);
\draw  [->] (p2) to (t1);
\draw  [->] (p1) to (t2);
\draw  [->] (p3) to (t2);
\draw  [->] (p2) to (t3);
\draw  [->] (p3) to (t3);

\node (q1) [place,tokens=1]  [right=\eofigdist of p1,label=above:$r_1$] {};
\node (q2) [place,tokens=1]  [right=\eodist of q1,label=above:$r_2$] {};
\node (q3) [place,tokens=1]  [right=\eodist of q2,label=above:$r_3$] {};
\node (s1) [transition] [below=\eodist of q1,label=left:$a$] {};
\node (s2) [transition] [below=\eodist of q3,label=right:$a$] {};

\draw  [->] (q1) to (s1);
\draw  [->] (q2) to (s1);
\draw  [->] (q2) to (s2);
\draw  [->] (q3) to (s2);

\node (b1) [place]  [right=\eofigdist of q1,label=above:$b_1$] {};
\node (b2) [place]  [right=\eodist of b1,label=above:$b_2$] {};
\node (b3) [place]  [right=\eodist of b2,label=above:$b_3$] {};
\node (u1) [transition] [below=\eodist of b2,label=left:$a$] {};

\draw  [->] (b1) to (u1);
\draw  [->] (b2) to (u1);

\end{tikzpicture}
\caption{Two non-cn-bisimilar markings, but with the same causal nets}
\label{cn-vs-icn-fig}
\end{figure}

Therefore, we want to relax the definition of causal-net bisimulation in order to equate the two nets discussed in the example above. 
The problem is essentially that causal-net bisimulation requires to fix the mappings of the current markings {\em before} the transition is selected, 
while it should be more correct to fix the mapping {\em after} the transition is selected, in order to work up to the 
choice of the isomorphism.
To achieve this, we have to rephrase the definition of causal-net bisimulation by exploiting 
 more relaxed definitions of folding and process.
 In particular, the {\em partial} folding below is defined as a folding  (cf. Definition \ref{folding-def}) 
 except that the mapping may be undefined, actually it is undefined only on the maximal conditions of the causal net; 
 this has the consequence that 
 (i) the mapping of the initial conditions 
 of the causal net is included in the initial marking of the net (of course, it coincides only if the mapping is 
 defined for all the initial conditions)
 and (ii) the post-set of each event has the same size of the post-set of the corresponding transition 
 (while the actual mapping of its post-set can be only partially included in the post-set of the corresponding transition).


\begin{defi}\label{p-folding-def}{\bf (Partial Folding and Partial Process)}
A {\em partial folding} from a causal P/T net $C = (B, L, E, \mathsf{m}_0)$ into a  P/T net system
$N(m_0) = (S, A, T, m_0)$ is a partial function $\rho: B \cup E \to S \cup T$, which is type-preserving, i.e., such that $\rho(B) \subseteq S$ and $\rho(E) \subseteq T$, satisfying the following:
\begin{enumerate}
\item $\forall b \in Max(C)$,  $\rho(b)$ is undefined, while $\forall b \not\in Max(C)$,  $\rho(b)$ is defined and $\forall e \in E$,  $\rho(e)$ is defined;
\item $L = A$ and $\mathsf{l}(e) = l(\rho(e))$ for all $e \in E$;
\item $|\mathsf{m}_0| = |m_0|$;
\item $\rho(\mathsf{m}_0) \subseteq m_0$, i.e., $ | \rho^{-1}(s) \cap \mathsf{m}_0 | \leq m_0(s)$;\footnote{To be precise, as $\rho$ is a partial function, by $\rho(\mathsf{m}_0) \subseteq m_0$ we mean that
if $\rho(\mathsf{m}_0)$ returns a marking, then that marking is contained in $m_0$. In general, if $B$ ia a set of conditions and $\rho$ is defined only on a subset $B' \subseteq B$,
then $\rho(B) = \rho(B')$; in case $B' = \emptyset$, then $\rho(B)$ is undefined. The same proviso applies also to the condition $ \rho(\post{e}) \subseteq \post{\rho(e)}$ in the last item of this definition.}
\item $\forall e \in E, \rho(\pre{e}) = \pre{\rho(e)}$, i.e., $\rho(\pre{e})(s) = | \rho^{-1}(s) \cap \pre{e} |$
for all $s \in S$;
\item $\forall e \in E$, $|\post{e}| = |\post{\rho(e)}|$ and $ \rho(\post{e}) \subseteq \post{\rho(e)}$, i.e., $| \rho^{-1}(s) \cap \post{e} | \leq \post{\rho({e})}(s)$ for all $s \in S$.
\end{enumerate}
A pair $(C, \rho)$, where $C$ is a causal net and $\rho$ a partial folding from  
$C$ to a net system $N(m_0)$, is a {\em partial process} of $N(m_0)$.
\end{defi}

Given a process $\pi = (C, \rho)$, it is possible to derive a unique partial process $\pi' = (C, \rho')$  --  where $\rho'$ is undefined
on $Max(C)$, while it is defined as $\rho$ on all the other conditions and on all the events of $C$ -- we call its {\em associated partial process}.

\begin{defi}\label{p-trans-process}{\bf (Moves of a partial process)}
Let $N(m_0) = (S, A, T, m_0)$ be a net system 
and let $(C_i, \rho_i)$, for $i = 1, 2$, be two partial processes of $N(m_0)$.
We say that $(C_1, \rho_1)$
moves in one step to $(C_2, \rho_2)$ through $e$, denoted by
$(C_1, \rho_1) \derivp{e} (C_2, \rho_2)$, if we have $C_1 [e\rangle C_2$
and $\rho_1 \subseteq \rho_2$.
\end{defi}

\begin{figure}[t]
    \centering
    
    \begin{tikzpicture}[
        every place/.style={draw,thick,inner sep=0pt,minimum size=6mm},
        every transition/.style={draw,thick,inner sep=0pt,minimum size=4mm},
        bend angle=30,
        pre/.style={<-,shorten <=1pt,>=stealth,semithick},
        post/.style={->,shorten >=1pt,>=stealth,semithick}    
    ]
    
    \node (a) [label=left:$N)\quad$]{};
    \node (p1) [place,tokens=1]  [label=above:$s_1$] {};   
    \node (p2) [place,tokens=1]  [right={1cm},label=above:$s_2$] {};   
    \node (t1) [transition] [below of = p2,label=right:$a$] {};
    \node (p3) [place] [below of = t1,label=right:$s_3$] {};
     \node (t2) [transition] [below of = p3,label=right:$b$] {};
    \node (p4) [place]  [below of = t2, label=below:$s_4$] {};
   
    \draw  [->] (p2) to (t1);
    \draw  [->] (t1) to (p3);
    \draw  [->] (p1) to (t2);
    \draw  [->] (p3) to (t2);
    \draw  [->] (t2) to (p4);
    
    \node (b) [right = {3cm} of p1, label=left:$C_0)$]{};
    \node (b1) [place]  [right = {0.1cm} of b, label=above:$b_1$] {};   
    \node (b2) [place] [right of = b1,label=above:$b_2$] {};

    \node (c) [right = {2.5cm} of b1, label=left:$C_1)$]{};
    \node (b11) [place]  [right = {0.1cm} of c, label=above:$b_1$] {};   
   \node (b12) [place]  [right = {1cm} of b11, label=above:$b_2$] {};   
    \node (e1) [transition] [below of = b12,label=left:$e_{a}$] {};
    \node (b21) [place] [below of = e1,label=left:$b_3$] {};
    
    \draw  [->] (b12) to (e1);
    \draw  [->] (e1) to (b21);

    \node (bp) [right = {2.8cm} of b11, label=left:$C_2)$]{};
    \node (b13) [place]  [right = {0.1cm} of bp, label=above:$b_1$] {};
    \node (b23) [place]  [right = {1cm} of b13, label=above:$b_2$] {};
    \node (e3) [transition] [below of = b23,label=left:$e_{a}$] {};
     \node (b33) [place]  [below of = e3, label=right:$b_3$] {};
   \node (e4) [transition] [below of = b33,label=right:$e_{b}$] {};
    \node (b43) [place]  [below of = e4, label=below:$b_4$] {};
 
    \draw  [->] (b23) to (e3);
    \draw  [->] (e3) to (b33);
    \draw  [->] (b13) to (e4);
    \draw  [->] (b33) to (e4);
    \draw  [->] (e4) to (b43);

    \end{tikzpicture}
    
    \caption{A finite marked P/T net $N$ and three causal nets}
    \label{fig:partial_example}
\end{figure}

\begin{exa}
Consider the marked net $N(s_1 \oplus s_2)$ and the three causal nets 
$C_0$, $C_1$ and $C_2$ in Figure \ref{fig:partial_example}, where by $e_a$ (or $e_b$) we mean 
the event with label $a$ (or $b$). The initial partial process for $N(s_1 \oplus s_2)$
is given by the pair $\pi_0 = (C_0, \emptyset)$, where the mapping is empty (or undefined); 
the only nontrivial condition that must be satisfied
is (3) of Definition \ref{p-folding-def}; hence, $\pi_0$ is a partial process, indeed. Let us denote by $a^n$
the transition $(s_2, a, s_3)$. Consider
the mapping $\rho_1 = \{b_2 \mapsto s_2, e_a \mapsto a^n\}$. It is easy to see that $\pi_1 = (C_1, \rho_1)$
is a partial process for $N(s_1 \oplus s_2)$, as also conditions (4), (5) and (6) of Definition \ref{p-folding-def} are satisfied.
Note also that $\pi_0 \derivp{e_a} \pi_1$ because $C_0[e_a\rangle C_1$ and $\emptyset \subseteq \rho_1$.
Now consider the mapping $\rho_2 = \{ b_1 \mapsto s_1, b_2 \mapsto s_2, b_3 \mapsto s_3, e_a \mapsto a^n, e_b \mapsto b^n\}$,
where by $b^n$ we denote the transition $(s_1 \oplus s_3, b, s_4)$. It is easy to see that $\pi_2 = (C_2, \rho_2)$
is a partial process for $N(s_1 \oplus s_2)$, as also conditions (4), (5) and (6) of Definition \ref{p-folding-def} are satisfied.
Note also that $\pi_1 \derivp{e_b} \pi_2$ because $C_1[e_b\rangle C_2$ and $\rho_1 \subseteq \rho_2$.
\end{exa}

The novel behavioral equivalence we propose is the following {\em i-causal-net bisimulation}, where 
the prefix $i-$ stands for {\em up to the isomorphism} of the current markings.

\begin{defi}\label{icn-bis-def}{\bf (i-causal-net bisimulation)} 
Let $N = (S, A, T)$ be a P/T net. An {\em i-causal-net bisimulation} (icn-bisimulation, for short)
is a relation $R$, composed of 
triples of the form $( \rho_1, C, \rho_2)$, where, for $i = 1, 2$, $(C, \rho_i)$ is a partial process of $N(m_{0i})$ for some $m_{0i}$,
such that if $( \rho_1, C, \rho_2) \in R$ then

\begin{itemize}[align=left]
\item[$i)$]
$\forall t_1, C', \rho_1'$ such that $(C, \rho_1) \derivp{e} (C', \rho'_1)$ with $\rho_1'(e) = t_1$, there exist $t_2,  \rho_2'$ such that
 $(C, \rho_2) \derivp{e} (C', \rho'_2)$, with $\rho_2'(e) = t_2$, and $( \rho'_1, C', \rho'_2) \in R$;

\item[$ii)$] and symmetrically, $\forall t_2, C', \rho_2'$ such that $(C, \rho_2) \derivp{e} (C', \rho'_2)$ with $\rho_2'(e) = t_2$, 
there exist $t_1,  \rho_1'$ such that 
$(C, \rho_1) \derivp{e} (C', \rho'_1)$, with $\rho_1'(e) = t_1$, and $( \rho'_1, C', \rho'_2) \in R$.
\end{itemize}

\noindent
Two markings $m_{1}$ and $m_{2}$ of $N$ are icn-bisimilar, 
denoted by $m_{1} \sim_{icn} m_{2}$, 
if there exists an i-causal-net bisimulation $R$ containing a triple of the form $( \rho^0,\, C^0,\, \rho^0)$, 
where:
\begin{itemize}
\item
$C^0$ contains no events, 
\item $\rho^0$ is undefined for all $b \in C^0$ (i.e., it is the empty function, also denoted by $\emptyset$) 
and, 
\item for $i = 1, 2$, $(C^0, \rho^0)$ is a partial process of $N(m_{i})$ for $m_{i}$ (i.e., this is the same as requiring that
$|Max(C^0)| = |m_{1}| = | m_{2}|$).
\qed
\end{itemize}
\end{defi}

Note that if $m_{1} \sim_{icn} m_{2}$, then $|m_{1}| = |m_{2}|$ because an i-casual-net bisimulation $R$ must 
contain the triple $( \rho^0, C^0, \rho^0)$ mentioned above. 
Moreover, whenever a triple $( \rho_1, C, \rho_2) \in R$ is reached by the icn-bisimulation game, we know that the current markings 
$m_1'$ and $m_2'$, reached from the initials
$m_{1}$ and $m_{2}$, respectively, must have the same size, equal to that of $Max(C)$.

Of course, $\sim_{icn}$ is an equivalence relation because the identity relation defined as $Id = \{( \rho, C, \rho) \mid (C, \rho)$ is 
a partial process of $N(m_{0})\}$ is an icn-bisimulation, the inverse relation $R^{-1}$ of an icn-bisimulation $R$ is an icn-bisimulation and, finally, the relational composition $R_1 \circ R_2$ of the 
icn-bisimulations $R_1$ and $R_2$ is an icn-bisimulation.
Moreover, $\sim_{cn}$ is finer than $\sim_{icn}$ because if $R_1$ is a causal-net bisimulation,
then $R_2 = \{(\rho'_1, C, \rho_2') \mid (\rho_1, C, \rho_2) \in R_1\}$, where for $i = 1, 2$, $(C, \rho'_i)$ is the partial process associated
to $(C, \rho_i)$, is an i-causal-net bisimulation.

\begin{exa}
Let us consider the net $N$ discussed in Example \ref{icn<cn-ex} (more precisely, $N$ is the union of the two nets, considered unmarked). 
By $a^s_l$ we denote the $a$-labeled transition with 
preset $s_1 \oplus s_2$, by $a^s_c$ that with preset $s_1 \oplus s_3$,
by $a^s_r$ that with preset $s_2 \oplus s_3$, by $a^r_l$ that with preset $r_1 \oplus r_2$ and, finally, by 
$a^r_r$ that with preset $r_2 \oplus r_3$.

Moreover, we denote by $C^0$ the causal net with no 
events and conditions $b_1,b_2, b_3$, while
we denote by $C^1$ the causal net extending $C^0$ with one $a$-labeled event $e_1$ such that $\pre{e_1}= b_1 \oplus b_2$ and
$\post{e_1}= \theta$, 
as depicted on the right of Figure \ref{cn-vs-icn-fig}. Similarly, we define $C^2$ as the extension of $C^0$  with one $a$-labeled 
event $e_2$ such that
$\pre{e_2}= b_1 \oplus b_3$ and $\post{e_2}= \theta$,  and also $C^3$ as the extension of $C^0$ with one $a$-labeled 
event $e_3$ 
such that $\pre{e_3}= b_2 \oplus b_3$ and $\post{e_3}= \theta$.

We can prove that $s_1 \oplus s_2 \oplus s_3 \sim_{icn} r_1 \oplus r_2 \oplus r_3$ as the following relation 

$
\begin{array}{rcl}
    R & = & \{ (\emptyset, C^0, \emptyset),\\
       & &  ((b_1 \mapsto s_1, b_2 \mapsto s_2, e_1 \mapsto a^s_l), C^1,  (b_1 \mapsto r_1, b_2 \mapsto r_2, e_1 \mapsto a^r_l)),\\
	&& ((b_1 \mapsto s_1, b_2 \mapsto s_3, e_1 \mapsto a^s_c), C^1,  (b_1 \mapsto r_1,  b_2 \mapsto r_2, e_1 \mapsto a^r_l)),\\
        & & ((b_1 \mapsto s_2, b_2 \mapsto s_3, e_1 \mapsto a^s_r), C^1,  (b_1 \mapsto r_2, b_2 \mapsto r_3, e_1 \mapsto a^r_r)),\\
      & &  ((b_1 \mapsto s_1, b_3 \mapsto s_2, e_2 \mapsto a^s_l), C^2,  (b_1 \mapsto r_1, b_3 \mapsto r_2, e_2 \mapsto a^r_l)),\\
%
	&& ((b_1 \mapsto s_1, b_3 \mapsto s_3, e_2 \mapsto a^s_c), C^2,  (b_1 \mapsto r_1,  b_3 \mapsto r_2, e_2 \mapsto a^r_l)),\\
        & & ((b_1 \mapsto s_2, b_3 \mapsto s_3, e_2 \mapsto a^s_r), C^2,  (b_1 \mapsto r_2, b_3 \mapsto r_3, e_2 \mapsto a^r_r)),\\
       & &  ((b_2 \mapsto s_1, b_3 \mapsto s_2, e_3 \mapsto a^s_l), C^3,  (b_2 \mapsto r_1, b_3 \mapsto r_2, e_3 \mapsto a^r_l)),\\
	&& ((b_2 \mapsto s_1, b_3 \mapsto s_3, e_3 \mapsto a^s_c), C^3,  (b_2 \mapsto r_1,  b_3 \mapsto r_2, e_3 \mapsto a^r_l)),\\
        & & ((b_2 \mapsto s_2, b_3 \mapsto s_3, e_3 \mapsto a^s_r), C^3,  (b_2 \mapsto r_2, b_3 \mapsto r_3, e_3 \mapsto a^r_r))     
             \},
\end{array}$

\noindent
is an icn-bisimulation
 containing a triple of the form $(\emptyset, C^0, \emptyset)$, 
where $(C^0, \emptyset)$ is a partial process of both $N(s_1\oplus s_2 \oplus s_3)$  and $N(r_1\oplus r_2 \oplus r_3)$.
Indeed, these two markings are icn-bisimilar, but not cn-bisimilar, because it is not possible to build a causal-net bisimulation by fixing
the initial isomorphism before choosing the matching transitions.
\end{exa}

As a final observation, we remark that, contrary to causal-net bisimulation, the definition of fully-concurrent bisimulation could be 
also rephrased in terms of partial processes, instead of (normal) processes, i.e., each triple $(\pi_1, f, \pi_2)$ in the relation 
is such that $\pi_1$ and $\pi_2$ are partial processes.
This because what is actually observed is only that $f$ is an isomorphism between 
$\mathsf{E}_{\mathsf{C}_1} = (E_1, \preceq_1)$ and $\mathsf{E}_{\mathsf{C}_2} = (E_2, \preceq_2)$, so that there is no need to fix
the mapping on the maximal conditions before the fully-concurrent bisimulation game starts.
In other words, also fully-concurrent bisimulation can be actually 
defined up to the choice of the isomorphism from maximal conditions of the current causal net to the tokens of the current marking.

We will prove that $\sim_{icn}$ is finer than $\sim_{fc}$ in Section \ref{icn-dec-sec} (cf. Theorem \ref{icn>fc-th}), as a byproduct of the 
alternative characterizations of i-causal-net bisimilarity and fully-concurrent bisimilarity in terms of OIMC bisimilarity 
and OIM bisimilarity, respectively.
This implication is strict, as illustrated by the following example.
 
\begin{exa}\label{ex-icn-vs-sfc}
Consider the nets 

$N = (\{s_1, s_2, s_3, s_4\},
\{a\}, \{(s_1 \oplus s_2, a, s_3 \oplus s_4)\})$ and $N' = (\{s_1', s_2', s_3'\}, \{a\}, \{(s_1', a, s_3')\})$.

\noindent
Of course, $s_1 \oplus s_2 \sim_{fc} s_1' \oplus s_2'$, as the 
generated partial orders are the same (and also the related markings have the same size), 
but $s_1 \oplus s_2 \nsim_{icn} s_1' \oplus s_2'$, as the generated causal nets are different.
\end{exa}

%
\section{Indexed Marking Semantics} \label{index-sec}
%
We define an alternative, novel token game semantics for Petri nets according to the {\em individual token
 philosophy}. A token is represented as an {\em indexed place}, i.e., as a pair $(s, i)$,
 where $s$ is the name of the place where the token is on, and $i$ is an index 
 assigned to the token such that different tokens 
 on the same place have different indexes. 
 In this way, a standard marking is turned into an {\em indexed} marking,
 i.e., a set of indexed places.

\begin{defi} \label{im} {\bf (Indexed marking)}
Given a finite net $N = (S,A,T)$, an {\em indexed marking} is a function $k: \; S \xrightarrow[]{} \mathcal{P}_{fin}(\nat)$ 
associating to each place a finite set of natural numbers, 
such that the associated (de-indexed) marking $m$ is obtained as $m(s) = \size{k(s)}$ for each $s \in S$. In this case, we write $\alpha(k) = m$. 
The support set $dom(k)$ is $\lbrace s \in S \mid k(s) \neq \emptyset \rbrace$.  
The set of all the indexed markings over $S$ is denoted by $\allim$.
An indexed place is a pair $(s, i)$ such that $s \in S$ and $i \in \nat$. A finite set of indexed places 
$\{ (s_1, i_1), \ldots , (s_n, i_n) \} \in \mathcal{P}_{fin}(S \times \nat)$
is also another way of describing an indexed marking.\footnote{Being a set, we are sure that  
$\not \exists j_1, j_2$ such that 
$s_{j_1} = s_{j_2} \; \wedge \; i_{j_1} = i_{j_2}$, i.e., each token on a place $s$ has an index
different from the index of any other token on $s$.}
Hence, $\allim = \mathcal{P}_{fin}(S \times \nat)$.
Each element of an indexed marking, i.e., each indexed place, is a token.

An indexed marking $k \in \allim$ is {\em closed} if $k(s) = \lbrace 1, 2, \ldots , \size{k(s)} \rbrace$ for all $s \in dom(k)$, i.e., there are no holes in the indexing.
If there exists a marked net $N(m_0)$ and 
a closed indexed marking $k_0$ such that $\alpha(k_0) = m_0$,
we say that $k_0$ is the {\em initial indexed} marking of $N$, and we write $N(k_0)$.
\end{defi}

Note that, given a marked net $N(m_0)$, the initial indexed marking $k_0$ is unique, because such $k_0$ is the only closed 
function from $S$ to $\mathcal{P}_{fin}(\nat)$ such that $\alpha(k_0) = m_0$.
However, it is interesting to observe that this modeling of the initial indexed marking is actually {\em up to isomorphism} of
the choice of the initial index assignment to multiple tokens on the same place.
For instance, if we have a marking composed of
two tokens on place $s$,
say $a$ and $b$ (to distinguish them), then both $\{(s_a, 1), (s_b, 2)\}$ and $\{(s_a, 2), (s_b, 1)\}$ are possible initial indexings.
However, this difference is completely inessential for the treatment that follows, as the two behavioral relations we study are defined 
up to isomorphism
of the chosen initial assignment. In fact, in the example above, this unique initial indexed marking is $\{(s, 1), (s, 2)\}$, that summarizes 
the two, more concrete marking representations above, up to isomorphism, but still giving individuality to each token for the future
by means of the index associated to the place. 

We define the difference between an indexed marking $k$ and a marking $m$ 
(such that $m(s) \leq |k(s)|$ for all $s \in S$) where for each $s$, $m(s)$ arbitrary tokens are removed from $k$ (hence, this operation
is nondeterministic)
as $\boxminus: \allim \xrightarrow[]{} \mathcal{M}(S) \xrightarrow[]{} \mathcal{P}(\allim)$
\begin{flalign*}
    &\hspace{2em}k \boxminus \theta = \lbrace k \rbrace &&\\
    &\hspace{2em}k \boxminus (s \oplus m) = (k \boxminus s) \boxminus m &&\\
    &\hspace{2em}\lbrace k_1 , \ldots k_n \rbrace \boxminus m = k_1 \boxminus m \cup \ldots \cup k_n \boxminus m &&\\
    &\hspace{2em}k \boxminus s = \lbrace k' \mid k'(s') =  k(s') \; \text{ if } s' \neq s , \text{ while }
                                    k'(s') = k(s) \setminus \lbrace n \rbrace \text{ if } s' = s \text{ and } n \in k(s) \rbrace &&
\end{flalign*}
\noindent 
and the deterministic operation of 
union of an indexed marking $k$ and a marking $m$ as $\boxplus: \allim \xrightarrow[]{} \mathcal{M}(S) \xrightarrow[]{} \allim$
\begin{flalign*}
    &\hspace{2em}k \boxplus \theta = k &&\\
    &\hspace{2em}k \boxplus (s \oplus m) = (k \boxplus s) \boxplus m &&\\
    &\hspace{2em}k \boxplus s = k' &&
\end{flalign*}
where for all $s' \in S \, , \; k'(s')$ is defined as:
\[   
k'(s') = 
     \begin{cases}
       k(s') &\quad\text{if } s' \neq s\\
       k(s) \cup \lbrace n \rbrace &\quad\text{if } s' = s, \; n = min(\nat \setminus k(s)) 
     \end{cases}
\]
where we use $min(H)$, with $H \in \mathcal{P}(\nat)$, to denote the least element of $H$.
Note that the difference between an indexed marking and a marking is a {\em set} of indexed markings: 
since it makes no sense to prefer a single possible execution over another, all possible choices for $n \in k(s)$ are to be considered.
The token game is modified accordingly, taking into account the individual token interpretation.

\begin{defi} \label{token-game-im}{\bf (Token game with indexed markings)}
Given a net $N = (S,A,T)$ and an indexed marking $k \in \allim$ such that $m = \alpha(k)$, we say that
a transition $t \in T$ is {\em enabled} at $k$ if  $\pre{t} \subseteq m$, denoted $k \imtrans{t}$.
If $t$ occurs, the firing of $t$ enabled at $k$ produces the indexed marking $k'$, denoted $k \imtrans{t} k'$, if 
\begin{itemize}
    \item[-] $\exists k'' \; \in \; k \boxminus \pre{t}$ and
    \item[-] $k' = k'' \boxplus \post{t}$. 
\end{itemize}
\end{defi}

Note that there can be more than one indexed marking produced by the firing of $t$,
but for all $k'$ such that $k \imtrans{t} k'$, it is true that $\alpha(k') = m \ominus \pre{t} \oplus{\post{t}}$. 

From now on, indexed markings will be always represented as sets of indexed places, i.e., 
we denote an indexed marking $k$ by  
$\{(s_1, n_1) \ldots (s_i, n_i) \}$ where $\size{k} = i$.

\begin{figure}[ht]
    \centering
    
    \begin{tikzpicture}[
        every place/.style={draw,thick,inner sep=0pt,minimum size=6mm},
        every transition/.style={draw,thick,inner sep=0pt,minimum size=4mm},
        bend angle=30,
        pre/.style={<-,shorten <=1pt,>=stealth,semithick},
        post/.style={->,shorten >=1pt,>=stealth,semithick}    
    ]
    
    \node (a) [label=left:$a)\qquad$]{};

    \node (p1) [place]  [label=above:$s_1$] {}
        [children are tokens]
        child {node [token] {1}};
    \node (t1) [transition] [below of = p1,label=left:$u$] {};
    \node (p2) [place] [below of = t1,label=left:$s_2$] {}
        [children are tokens]
        child {node [token] {1}}
        child {node [token, fill=red] {2}}
        child {node [token] {3}};
    
    \node (t2) [transition] [below of = p2,label=left:$v$] {};
    \node (p3) [place, tokens=0] [below of = t2,label=left:$s_3$] {};

    \draw  [->] (p1) to (t1);
    \draw  [->] (t1) to node[auto,swap] {2} (p2);
    \draw  [->] (p2) to (t2);
    \draw  [->] (t2) to (p3);
    
    
    \node (p4) [place]  [right = {3.8cm} of p1, label=above:$s_1$] {}
        [children are tokens]
        child {node [token, fill=red] {1}};
    \node (t4) [transition] [below of = p4,label=left:$u$] {};
    \node (p5) [place] [below of = t4,label=left:$s_2$] {}
        [children are tokens]
        child {node [token] {1}}
        child {node [token] {3}};
    \node (t5) [transition] [below of = p5,label=left:$v$] {};
    \node (p6) [place, tokens=1] [below of = t5,label=left:$s_3$] {}
        [children are tokens]
        child {node [token, fill = blue] {1}};

    \draw  [->] (p4) to (t4);
    \draw  [->] (t4) to node[auto,swap] {2} (p5);
    \draw  [->] (p5) to (t5);
    \draw  [->] (t5) to (p6);

    \node (b)  [right={3.7cm} of a,label=left:$b)\quad$]{};


    \node (p7) [place]  [right = {3.8cm} of p4, label=above:$s_1$] {};
    \node (t7) [transition] [below of = p7,label=left:$u$] {};
    \node (p8) [place] [below of = t7,label=left:$s_2$] {}
        [children are tokens]
        child {node [token] {1}}
        child {node [token, fill = blue] {2}}
        child {node [token] {3}}        
        child {node [token, fill = blue] {4}};
    \node (t8) [transition] [below of = p8,label=left:$v$] {};
    \node (p9) [place] [below of = t8,label=left:$s_3$] {}
        [children are tokens]
        child {node [token] {1}};

    \draw  [->] (p7) to (t7);
    \draw  [->] (t7) to node[auto,swap] {2} (p8);
    \draw  [->] (p8) to (t8);
    \draw  [->] (t8) to (p9);

    \node (c)  [right={3.8cm} of b,label=left:$c)\; $]{};

    \end{tikzpicture}
    
    \caption{Execution of the transition labeled by $v$, then of the transition labeled by $u$, on a net with initial marking 
    $m_0 = s_1 \oplus 3s_2$. Tokens to be consumed are in red, generated ones in blue.}
    \label{fig:token-game-1}
\end{figure}

\begin{exa}\label{token-game-im-example}
In Figure \ref{fig:token-game-1}(a) a simple marked net $N$ is given. 
The initial marking is $m_0 = s_1 \oplus 3s_2$, and it is not difficult to see that 
the net system $N(m_0)$ is 5-bounded.
The initial indexed marking is $k_0 = \{ (s_1, 1), (s_2, 1), (s_2, 2), (s_2, 3) \}$.

\noindent
Let us suppose that transition $t_2$, labeled by $v$, occurs.
There are three possible ways to remove a token from $s_2$: 
removing $(s_2, 1)$, or removing $(s_2, 2)$, or removing $(s_2, 3)$.
Indeed, the operation $k_0 \boxminus \pre{t_2}$ yields 
a set of three possible indexed markings,
each one a possible result of the difference:
$\{                                      
   \{ (s_1, 1), (s_2, 2), (s_2, 3) \},
   \{ (s_1, 1), (s_2, 1), (s_2, 3) \},
   \{ (s_1, 1), (s_2, 1), (s_2, 2) \}
\}$.
Let us choose, for the sake of the argument, 
that the token deleted by $t_2$ is $(s_2, 2)$,
i.e. choose $k' = \{ (s_1, 1), (s_2, 1), (s_2, 3) \}$.
The union $k' \boxplus \post{t_2}$ easily yields the 
indexed marking $k_1 = \{ (s_1, 1), (s_2, 1), (s_2, 3), (s_3, 1) \}$,
as depicted in Figure \ref{fig:token-game-1}(b).
Note that the choice of $k'$ was arbitrary and two other values of $k_1$ are possible.
Indeed, from Definition \ref{token-game-im}, we know that the transition relation
on indexed markings is nondeterministic.
However, the resulting marked net is the same for all three cases, that is, 
the same of Figure \ref{fig:token-game-1}(b) without indexes.

\noindent
Now we suppose that (given the indexed marking $k_1$ above) transition $t_1$, labeled by $u$, occurs.
In that case, $k_1 \boxminus \pre{t_1}$ yields the singleton set 
$\{                                      
   \{ (s_2, 1), (s_2, 3), (s_3, 1) \}
\}$ of indexed markings,
and so we take $k'' = \{ (s_2, 1), (s_2, 3), (s_3, 1) \}$.
Since $\post{t_1} = s_2 \oplus s_2$, we show in detail how $k'' \boxplus \post{t_1}$ is computed.
First, we apply the definition for union with non-singleton multisets: 
$k'' \boxplus (s_2 \oplus s_2) = (k'' \boxplus s_2) \boxplus s_2$.
Then, we compute $k'' \boxplus s_2$: since the least free index for the place $s_2$ is 2,
$k'' \boxplus s_2 = \{ (s_2, 1), (s_2, 2) (s_2, 3), (s_3, 1) \}$.
Now we apply again the definition: 
note that this time the least free index for $s_2$ is 4,
and the final result is $k_2 = \{ (s_2, 1), (s_2, 2) (s_2, 3), (s_2, 4), (s_3, 1) \}$.
The resulting marked net is depicted in Figure \ref{fig:token-game-1}(c).
\end{exa}
The notation for tokens in the token game has become less intuitive, 
so in Table \ref{table-tokens} we provide a 
comparison between the one used in the previous sections and 
the one we will use in the following part of this work.
Given a transition $t$ such that $k \imtrans{t} k'$
and $m \trns{t} m'$, where $\alpha(k) = m$ and $\alpha(k') = m'$,
assume $k'' \in k \boxminus \pre{t}$
such that $k' = k'' \boxplus \post{t}$.

\begin{table}[!h]
\begin{tabular}{ccccc}\toprule
 & & generated & deleted & untouched \\\midrule
 & $m \trns{t} m'$ & $\post{t}$ & $\pre{t}$ & $m \ominus \pre{t}$ \\\hline
 & $k \imtrans{t} k'$ & $\generated{}$ & $\deleted{}$ & $\old{}$ \\\bottomrule
\end{tabular}
\caption{Different notation for tokens in the token game. On the first line, the collective case.
On the last one, the individual case.}
\label{table-tokens}
\end{table}

\begin{defi}\label{firing-sequence-im}{\bf (Firing sequence with IM)}
Given a finite net $N = ( S, A, T)$ and an indexed marking $k$, 
a {\em firing sequence} starting at $k$ is defined inductively as follows:
\begin{itemize}
\item $k\llbracket \epsilon\rangle k$ is a firing sequence (where $\epsilon$ denotes an empty sequence of transitions) and
\item if $k\llbracket \sigma\rangle k'$ is a firing sequence and $k' \llbracket t\rangle k''$, then
$k\llbracket \sigma t\rangle k''$ is a firing sequence. 
\end{itemize}

\noindent
The set of reachable indexed markings from $k$ is $\imtrans{k} = 
\{ k' \mid \exists \sigma \, . \, k \imtrans{\sigma} k'\}$. 
Given a net $N(k_0)$ with $k_0$ an initial indexed marking, we call 
$IM(N(k_0))$ the set of reachable indexed marking of $N(k_0)$.
When the initial indexed marking $k_0$ is clear from the context, we may just write $IM(N)$.
\end{defi}

\begin{prop} \label{reachable-im-finite}
Given a finite bounded net $N = (S,A,T, m_0)$, 
the set $IM(N) \; \subseteq \allim$ of reachable indexed markings is finite.
\proof
Since $N(m_0)$ is bounded, 
  an index $h \in \nat$ exists such that the net is $h$-bounded.
The initial indexed marking $k_0$, with $\alpha(k_0) = m_0$, being closed, is such that 
no indexed places in $k_0$ has an index larger than $h$. 
Each token 
in a reachable indexed marking $k$ has always index less than, or equal to $h$, because the net is $h$-bounded 
and, by definition of 
$\boxplus$, in the token game we choose always the least available index for a newly produced token.
Therefore, $IM(N)$ is finite because $IM(N) \subseteq \mathcal{P}_{fin}(S \times \{1, \ldots , h\})$, which is finite as $S$ is finite as well.
\qed
\end{prop}

\section{Ordered Indexed Marking Semantics}\label{oim-sec}
%

Vogler \cite{Vog91} introduces {\it ordered markings} (OM for short) 
to describe the state of a safe marked net. 
An ordered marking consists of a safe marking together with a preorder which 
reflects precedences in the generation of tokens.
This is reflected in the token game for OM: if $s$ precedes some $s''$ 
in the old OM and $s''$ is used to produce a new token $s'$, 
then $s$ must precede $s'$ in the new OM. 
The key idea of Vogler's decidability proof for safe nets is that the OM
obtained by a sequence of transitions of a net is the same as the one
induced by a process, whose events correspond to that sequence of transitions, 
on the net itself. Vogler
defines OM bisimulation and shows that it coincides
with fully-concurrent bisimulation. Since ordered markings are finite objects and the reachable ordered markings are finitely many,
the candidate relations to be OM bisimulations are finite and finitely many, so that OM bisimilarity is decidable.
He himself hinted at a possibility \cite{Vog91} of extending
the result to bounded nets, but suggested 
that it would have been technically quite involved \cite{Vog91}(p. 503).

We adapt his approach by defining
a semantics based on {\em ordered indexed markings}, 
taking into account the individual token interpretation of nets,
and proving that an extension to bounded nets is indeed possible.

\begin{defi} \label{oim} {\bf (Ordered indexed marking)}
Given a P/T net $N = (S,A,T)$ and an indexed marking $k \in \allim$, the pair $(k, \leq)$ is an ordered 
indexed marking if $\leq \subseteq k \times k$ is a preorder, i.e. a reflexive and transitive relation.
The set of all possible ordered indexed markings of $N$ is denoted by $OIM(N)$.

If $k_0$ is the initial indexed marking of $N$, 
we define the {\em initial ordered indexed marking}, denoted by 
$init(N(k_0))$, as $(k_0, k_0 \times k_0)$, where the initial preorder relates each token with each other one.
If the initial indexed marking $k_0$ is clear from the context, 
we write simply $init(N)$ to denote the initial ordered indexed marking.
\end{defi}
 
\begin{defi} \label{token-game-oim}{\bf (Token game with ordered indexed markings)}
Given a P/T net $N = (S,A,T)$ and an ordered indexed marking $(k, \leq)$, we say that 
a transition $t \in T$ is {\em enabled} at $\oim{}$ if $k \imtrans{t}$;
this is denoted by $\oim{} \imtrans{t}$.
The firing of $t$ enabled at $\oim{}$ may produce an ordered indexed marking $\oimp{}$ -- 
and we denote this by $\oim{} \imtrans{t} \oimp{}$  -- where:
\begin{itemize}
    \item $\exists k'' \in k \boxminus \pre{t}$ and  $k' = k'' \boxplus \post{t}$, and
    
    \item for all $(s_h, i_h), (s_j, i_j) \in k' \; , \; (s_h, i_h) \leq' (s_j, i_j)$ if and only if:
        \begin{enumerate}
            \item $(s_h, i_h), (s_j, i_j) \in k''$ (i.e., the two tokens belong to the untouched part of the indexed marking) and $(s_h, i_h)\leq (s_j, i_j)$, or
            \item $(s_h, i_h), (s_j, i_j) \in k' \setminus k''$ (i.e., the two tokens are generated by the firing), or
            \item $(s_h, i_h) \in k'' , \; (s_j, i_j) \in k' \setminus k''$ 
                and $\exists (s_l, i_l) \in k \setminus k''$ (i.e., $(s_l, i_l)$ is a token consumed by the firing of $t$) such that
                $(s_h, i_h)\leq (s_l, i_l)$. 
        \end{enumerate}
\end{itemize}  
\end{defi}

Note that, as for indexed markings, many different ordered indexed markings are produced from the firing of $t$. This means that also the transition relation for ordered indexed markings is nondeterministic.
Moreover, in the same fashion as Vogler's work \cite{Vog91}, 
the preorder reflects the precedence in the generation of tokens,
which is not strict, i.e., if tokens $(s_1, n_1)$ and $(s_2, n_2)$ are generated
together (case (2) above) we have both $(s_1, n_1) \leq (s_2, n_2)$ and $(s_2, n_2) \leq (s_1, n_1)$.

\begin{exa}\label{token-game-oim-example}
Consider again the net in Figure \ref{fig:token-game-1}
and the first part of the execution of Example \ref{token-game-im-example},
i.e., $k_0 \imtrans{t_2} k_1$, where $k_0 = \{ (s_1, 1), (s_2, 1), (s_2, 2), (s_2, 3) \}$ and $k_1 = \{ (s_1, 1), (s_2, 1), $ \\
$(s_2, 3), (s_3, 1) \}$.
According to Definition \ref{oim}, the initial ordered indexed marking is $(k_0, \leq_0)$,
where $\leq_0 = k_0 \times k_0$. 
When $t_2$ fires, token $(s_2, 2)$ is removed and token $(s_3, 1)$ is generated,
while all other tokens are untouched. 
Let us denote the preorder induced by the firing of $t_2$ as $\leq_1$.
According to item 2 of Definition \ref{token-game-oim},
since $(s_3, 1)$ is generated by the firing of $t_2$, we have $(s_3,1) \leq_ 1 (s_3, 1)$.
According to item 1 of Definition \ref{token-game-oim}, 
the preorder on all tokens untouched by $t_2$ remains the same, 
therefore, e.g., $(s_2, 3) \leq_1 (s_1,1)$ and viceversa.
Furthermore, consider $(s_1, 1)$ and $(s_3,1)$: we have that $t_2$ generates $(s_3, 1)$, 
deletes $(s_2, 2)$ and leaves $(s_1, 1)$ untouched. Since $(s_1, 1) \leq_0 (s_2, 2)$, 
by  item 3 of Definition \ref{token-game-oim} we have $(s_1, 1) \leq_1 (s_3, 1)$.
The same reasoning applies to all untouched tokens.
Summing up, we have $(k_0, \leq_0) \imtrans{t_2} (k_1, \leq_1)$ where 
$\leq_1 \; =\;  \leq_0 \setminus 
\{ ((s_i, n_i), (s_j, n_j)) \in k_0 \mid (s_i, n_i) = (s_2, 2) \lor (s_j, n_j) = (s_2, 2) \} \; \cup \;
\{
((s_1, 1), (s_3, 1)),
((s_2, 1), (s_3, 1)),
((s_2, 3), (s_3, 1)), 
((s_3, 1), (s_3, 1))
\}
$.
\end{exa}

\begin{defi}\label{firing-sequence-oim}{\bf (Firing sequence with OIM)}
A {\em firing sequence} starting at $\oim{}$ is defined inductively as follows:
\begin{itemize}
\item $\oim{} \imtrans{\epsilon} \oim{} $ is a firing sequence (where $\epsilon$ denotes an empty sequence of transitions) and
\item if $\oim{} \imtrans{\sigma} \oimp{}$ is a firing sequence and $\oimp{}' \imtrans{t} \oims{}$, then 
$\oim{} \imtrans{\sigma t} \oims{}$ is a firing sequence. 
\end{itemize}

\noindent
The set of reachable ordered indexed markings from $\oim{}$ is 

$\imtrans{\oim{}} = 
\{ \oimp{} \mid \exists \sigma \, . \, \oim{} \imtrans{\sigma} \oimp{}\}$. 

\noindent
Given an initial indexed marking $k_0$, 
the set of all the reachable ordered indexed markings of $N(k_0)$ (starting from $init(N(k_0)) = (k_0, k_0 \times k_0)$)
is denoted by $\imtrans{init(N)}$.
\end{defi}

\begin{prop} \label{reachable-oim-finite}
Given a bounded net $N = (S,A,T, m_0)$, $\imtrans{init(N)}$ is finite.

\proof
The set $IM(N)$  of reachable indexed markings is finite
by Proposition \ref{reachable-im-finite}.
The set of possible preorders for an indexed marking
$k = \{ (s_1, n_1) \ldots (s_j, n_j) \} \in IM(N)$ 
is finite, because $\leq \subseteq k \times k$. 
Therefore, $\imtrans{init(N)}$ is finite.
\qed
\end{prop}

\subsection{Ordered indexed marking and causality-based semantics}

If $\pi = (C, \rho)$ is a process of a marked net $N(m_0)$ 
and $k_0$ is the initial indexed marking for $N(m_0)$ (i.e. $\alpha(k_0) = m_0$ and $k_0$ is closed), 
we also say that $\pi$ is a process of $N(k_0)$.
Given a firing sequence of a net $N(k_0)$,
there is an operational preorder on tokens obtained by Definition \ref{firing-sequence-oim},
and a preorder $\leq_\pi$ derived from the causal net $C$ of a process $\pi$ 
which models that execution.

In order to relate the execution of an event sequence $\sigma$ of $\pi$
and its corresponding firing sequence $\rho(\sigma)$ on the actual net $N(k_0)$, one must define how
maximal conditions in the process $\pi$ are mapped to indexed places. 
Indeed, the firing sequence $\rho(\sigma)$ on $N(k_0)$, corresponding to the execution of an event sequence $\sigma$ of $\pi$,
might generate many ordered indexed markings, depending on the choice of the initial mapping from conditions 
to tokens of the initial markings
as well as on the choice of the mapping from newly generated conditions of a new event $e$
to multiple tokens on the same place generated by the corresponding transition $\rho(e)$,
as illustrated in the following Example \ref{pi_but_not_oim}.
To reconcile abstract process semantics and concrete indexed marking semantics, 
we inductively define a \textit{process sequence} for $\pi$, which contains both
an event sequence $\sigma$ for $\pi$ and 
a mapping $\delta$ from maximal conditions of the process $\pi$ to indexed places of the current marking of the net. 

In the following, we may denote an indexed place $(s,i)$ as $p$ when it is not needed to make place $s$ and index $i$ explicit.

\begin{defi}\label{proc-seq}{\bf (Process sequence for a process)}\label{proc-seq-def}
Given a marked net $N(k_0)$ and $\leq_0 = k_0 \times k_0$, a \emph{process sequence} is inductively defined as follows:
\begin{itemize}
    \item $(k_0, \leq_0) \llbracket \epsilon, \delta_0 |\rangle (k_0, \leq_0)$ 
    is a process sequence for $\pi_0 = (C^0, \rho_0)$, 
    i.e., a process of $N(k_0)$ with empty set of events,
    where $\epsilon$ denotes the empty event sequence and 
    $\delta_0$ is a bijective mapping between $Max(C^0)$ and $k_0$ such that for each $b \in Max(C^0)$, 
    $\delta_0(b) = (\rho_0(b), i)$ for some
    $(\rho_0(b), i) \in k_0$;
    \item if $(k_0, \leq_0) \llbracket \sigma, \delta |\rangle (k, \leq)$ 
    is a process sequence for $\pi = (C, \rho)$ and
    $\pi \deriv{e} \pi' = (C', \rho')$, then we have that
    $(k_0, \leq_0) \llbracket \sigma e, \delta' |\rangle (k', \leq')$
    is a process sequence for $\pi'$, where:
    \begin{itemize}[-]
        \item Let $k'' = k \setminus \delta(\pre{e})$.\footnote{By abuse of notation, we write 
        $\delta(\pre{e})$ for the set $\{ \delta(b) \mid b \in \pre{e}  \}$.} Then, $k' = k'' \boxplus \post{\rho'(e)}$.
        \item $\delta'$ is a bijective mapping between $Max(C')$ and $k'$ defined as $\delta'(b) = \delta(b)$ if $b \in Max(C)$, while 
        on $Max(C') \setminus Max(C) = \post{e}$, $\delta'$ is a map from $\post{e}$ to $k' \setminus k'' = \{(\rho’(b), i) \mid b \in \post{e}$ $ \wedge (\rho’(b), i ) \in k’\}$ such that $\delta'(b) = (\rho'(b), i)$ for some $(\rho'(b), i ) \in k'$.
        \item For all $p_1, p_2 \in k'$, $p_1 \leq' p_2$ if and only if
            \begin{enumerate}
               \item $p_1, p_2 \in k \setminus \delta(\pre{e})$ and $p_1\leq p_2$, or
                \item $p_1, p_2 \in \delta'(\post{e})$, or
                \item $p_1 \in k \setminus \delta(\pre{e})$ ,  $p_2 \in \delta'(\post{e})$ 
                and $\exists p_1' \in \delta(\pre{e})$ such that $p_1\leq p_1'$. 
            \end{enumerate}
    \end{itemize}
\end{itemize}
\end{defi}

\begin{rem}{\bf (Process sequence for a partial process)}
Note that the definition of process sequence can be defined also w.r.t. {\em partial} processes. In fact, given $\leq_0 = k_0 \times k_0$,

\begin{itemize}
    \item $(k_0, \leq_0) \llbracket \epsilon, \delta_0 |\rangle (k_0, \leq_0)$ 
    is a process sequence for the {\em partial} process $\pi_0 = (C^0, \rho_0)$, 
    i.e., a partial process of $N(k_0)$ with empty set of events,
    where $\epsilon$ denotes the empty event sequence and 
    $\delta_0$ is a bijective mapping between $Max(C^0)$ and $k_0$;
    \item if $(k_0, \leq_0) \llbracket \sigma, \delta |\rangle (k, \leq)$ 
    is a process sequence for the {\em partial} process $\pi = (C, \rho)$ and
    $\pi \derivp{e} \pi' = (C', \rho')$, then we have that
    $(k_0, \leq_0) \llbracket \sigma e, \delta' |\rangle (k', \leq')$
    is a process sequence for $\pi'$, where:
    \begin{itemize}[-]
        \item Let $k'' = k \setminus \delta(\pre{e})$. Then, $k' = k'' \boxplus  \post{\rho'(e)}$.
        \item $\delta'$ is a bijective mapping between $Max(C')$ and $k'$ defined as $\delta'(b) = \delta(b)$ if $b \in Max(C)$, while 
        on $Max(C') \setminus Max(C) = \post{e}$, $\delta'$ is a map from $\post{e}$ to the set $k' \setminus k''$.
        \item For all $p_1, p_2 \in k'$, $p_1 \leq' p_2$ is defined as above.
            \end{itemize}
\end{itemize}
Hence, in the following, we simply say that $(k_0, \leq_0) \llbracket \sigma, \delta |\rangle (k, \leq)$ 
is a process sequence for $\pi = (C, \rho)$, where the fact that $\pi$ is a process or a partial process is almost irrelevant.
In particular, in Section \ref{fc-dec-sec} we will use this definition w.r.t. processes, while in Section \ref{icn-dec-sec} w.r.t. partial processes.
\end{rem}

The definition of process sequence (w.r.t. processes or partial processes) is nondeterministic. 
First, the initial step allows for different choices
for the initial $\delta_0$ function, which is a concrete mapping from minimal conditions to tokens of the initial indexed marking $k_0$
(respecting the function $\rho_0$ in case we consider processes instead of partial processes); and then, in the 
inductive case, $\delta'$ can be any bijection from the newly generated conditions
to the newly generated tokens in $k'$ (respecting $\rho'$, in case we consider processes).
However, this kind of nondeterminism is just apparent, because
$\delta$ works on the concrete net and,
by the indexes chosen initially for each minimal condition as well as for the new (i.e., generated) tokens in the same place, 
different concrete runs (i.e., with a different indexing of tokens due to a different choice of $\delta$) may originate isomorphic processes only,
as illustrated by the following example.

\begin{figure}[ht]
    \centering
    
    \begin{tikzpicture}[
        every place/.style={draw,thick,inner sep=0pt,minimum size=6mm},
        every transition/.style={draw,thick,inner sep=0pt,minimum size=4mm},
        bend angle=30,
        pre/.style={<-,shorten <=1pt,>=stealth,semithick},
        post/.style={->,shorten >=1pt,>=stealth,semithick}    
    ]
    
    \node (ap) [label=right:$a)$]{};
    \node (b1) [place]  [right={1cm} of ap, label=above:$b_1$] {};   
    \node (b2) [place]  [right of = b1, label=above:$b_2^1$] {};
    \node (b3) [place]  [right of  = b2, label=above:$b_2^2$] {};
    \node (b4) [place]  [right of = b3, label=above:$b_2^3$] {};
    \node (e1) [transition] [below of = b3, label=right:$e_v$] {};
    \node (b5) [place]  [below of = e1, label=below:$b_3$] {};
    \node (e2) [transition] [below of = b1, label=right:$e_u$] {};
    \node (b6) [place]  [below left of = e2, label=below:$b_2^4$] {};
    \node (b7) [place]  [below right of = e2, label=below:$b_2^5$] {};

    \draw  [->] (b3) to (e1);
    \draw  [->] (e1) to (b5);
    \draw  [->] (b1) to (e2);
    \draw  [->] (e2) to (b6);
    \draw  [->] (e2) to (b7);
        
    \node (b)  [right={6.5cm} of ap,label=left:$b)\quad$]{};

    \node (p4) [place]  [right={0cm} of b, label=above:$s_1$] {};
    \node (t4) [transition] [below of = p4,label=left:$u$] {};
    \node (p5) [place] [below of = t4,label=left:$s_2$] {}
        [children are tokens]
        child {node [token] {1}}
        child {node [token] {3}}
        child {node [token] {4}}
        child {node [token] {5}};
    \node (t5) [transition] [below of = p5,label=left:$v$] {};
    \node (p6) [place, tokens=1] [below of = t5,label=left:$s_3$] {}
        [children are tokens]
        child {node [token] {1}};

    \draw  [->] (p4) to (t4);
    \draw  [->] (t4) to node[auto,swap] {2} (p5);
    \draw  [->] (p5) to (t5);
    \draw  [->] (t5) to (p6);
    
    \node (c)  [right={3cm} of b,label=left:$c)\quad$]{};

    \node (p7) [place]  [right = {0cm} of c, label=above:$s_1$] {};
    \node (t7) [transition] [below of = p7,label=left:$u$] {};
    \node (p8) [place] [below of = t7,label=left:$s_2$] {}
        [children are tokens]
        child {node [token] {1}}
        child {node [token] {2}}        
        child {node [token] {3}}
        child {node [token] {4}};
    \node (t8) [transition] [below of = p8,label=left:$v$] {};
    \node (p9) [place] [below of = t8,label=left:$s_3$] {}
        [children are tokens]
        child {node [token] {1}};

    \draw  [->] (p7) to (t7);
    \draw  [->] (t7) to node[auto,swap] {2} (p8);
    \draw  [->] (p8) to (t8);
    \draw  [->] (t8) to (p9);
    
    \end{tikzpicture}
    
    \caption{The causal net of a process and two possible resulting indexed markings starting from the net of Figure \ref{fig:token-game-1}(a).}
    \label{fig:proc-seq}
\end{figure}

\begin{exa}\label{pi_but_not_oim}
Consider the net in Figure \ref{fig:token-game-1}(a), with $k_0 = \{ (s_1, 1), (s_2, 1), (s_2, 2), (s_2, 3) \}$, 
and assume that the initial $\delta_0$ maps $b_1$ to $s_1$, and $b^i_2$  to
$(s_2, i)$ for $i = 1, 2, 3$.
Consider the transition sequence $t_1 \, t_2$ (with label $u \, v$).
Transition $t_1$ consumes the token $(s_1, 1)$ and generates tokens $(s_2, 4)$ and $(s_2, 5)$; 
assume that transition $t_2$ consumes the token $(s_2, 2)$ and generates the token $(s_3, 1)$. 
Given that there is no causality between the firing of the two transitions, 
the process $\pi$, whose causal net is in Figure \ref{fig:proc-seq}(a), models that execution, with trivial mappings
(i.e., the mapping $\delta$ after the execution of the two transitions maps $b^i_2$ to
$(s_2, i)$ for $i = 1, 3, 4, 5$ and $b_3$ to $(s_3, 1)$).
Let $\sigma = e_u e_v$ be the event sequence of the process $\pi$, such that $\rho(\sigma) = t_1 \, t_2$.

Consider now to swap the execution order of these two independent transitions, so that the transition sequence 
now is $t_2 \, t_1$ (with label $v \, u$), where $t_2$ consumes 
the token $(s_2, 2)$ and generates the token $(s_3, 1)$, while $t_1$ consumes the token $(s_1, 1)$ and generates the 
tokens $(s_2, 2)$ and $(s_2, 4)$.
Again, there is no causality between them and so 
the process, whose causal net is in Figure \ref{fig:proc-seq}(a), models that execution, with trivial mappings albeit different from the first one
(in particular, the mapping $\delta'$ after the execution of the transition sequence $t_2 \, t_1$ is such that
$\delta'(b^4_2) = (s_2, 2)$ and $\delta'(b^5_2) = (s_2, 4)$).
Indeed, the two processes are isomorphic
(but the latter with a different event sequence $\sigma' = e_v e_u$ such that $\rho(\sigma')  = t_2 \, t_1$).
However, the resulting indexed markings (cf. Figure \ref{fig:proc-seq}(b) and (c)) are different.
As a matter of fact, the process sequence which gives origin to the net in Figure \ref{fig:proc-seq}(b) ends with a mapping $\delta$ whose image is the set $\{ (s_2, 1), (s_2, 3), (s_2, 4), (s_2, 5), (s_3, 1) \}$, 
while the one related to Figure \ref{fig:proc-seq}(c), i.e., $\delta'$, has the set  $\{ (s_2, 1), (s_2, 2) (s_2, 3), (s_2, 4), (s_3, 1) \}$ as image. 

Indeed, it is not even enough to keep track of the order of events to 
compute the resulting indexed markings: 
one must also consider how conditions are mapped to individual tokens 
initially. 
In fact, if we consider a different initial $\delta_0'$ (compatible with the initial $\rho_0$, that is not modified, if we consider 
processes instead of partial processes) 
mapping $b^3_2$ on $(s_2, 2)$  and $b^2_2$ on $(s_2, 3)$, and 
the same transition sequence $t_1 \, t_2$ above, where, however, the second transition consumes the token $(s_2, 3)$ 
and generates the token $(s_3, 1)$,
we get the same process $\pi$ of Figure \ref{fig:proc-seq}(a) (with the same event 
sequence $\sigma$ on $\pi$), but the final indexed marking is 
$\{ (s_2, 1), (s_2, 2) (s_2, 4), (s_2, 5), (s_3, 1) \}$.

Of course, there is also an instance of the transition sequence $t_1 \, t_2$ (with label $u \, v$) of the net, where $t_2$ consumes 
one of the tokens produced by $t_1$, but in this case the transition sequence of the
process (and the process itself) would be different. However, note that after firing $t_1$, the resulting indexed marking is 
$\{ (s_2, 1), (s_2, 2) (s_2, 3), (s_2, 4), (s_2, 5) \}$.
Now, $\delta'$ can be chosen in such a way that $b^4_2$ is mapped to $(s_2, 4)$ and $b^5_2$ is mapped to $(s_2, 5)$. If $t_2$ consumes
the token $(s_2, 4)$, then the reached indexed marking is $\{ (s_2, 1), (s_2, 2) (s_2, 3), (s_2, 5), (s_3, 1) \}$. However, the same process
can be obtained by choosing $\delta'$ in such a way that $b^4_2$ is mapped to $(s_2, 5)$ and $b^5_2$ is mapped to $(s_2, 4)$, with $t_2$ that consumes
token $(s_2, 5)$, but in such a case the resulting indexed marking is $\{ (s_2, 1), (s_2, 2) (s_2, 3), (s_2, 4), (s_3, 1) \}$.
\end{exa}

\begin{lem} \label{leq-minimal} {\bf (A minimality condition for $\leq$)}
Let $\pi = (C, \rho)$ be a process of $N(k_0)$ and, moreover, let
$(k_0, \leq_0) \llbracket \sigma, \delta |\rangle (k, \leq)$ be a process sequence for $\pi$.
For all $b \in Max(C)$ (i.e., for all $b$ such that $\delta(b) \in k$),
if $b \in Min(C)$ then:
\begin{itemize}
    \item $\delta(b) \in k_0$, and
    \item for all $b' \in Max(C)$, we have 
    $\delta(b) \leq \delta(b')$.
\end{itemize}

\proof
By induction on the length of $\sigma$. 
\qed
\end{lem}
\noindent
In other words, if $b \in Max(C)$ and also $b \in Min(C)$, then the current token $(\rho(b), i) \in k$ was actually already present 
in the initial indexed marking $k_0$ and 
it is also minimal for the preorder $\leq$. 
Note that the lemma above holds even in case $\pi = (C, \rho)$ is a {\em partial} process of $N(k_0)$.

\begin{thm} \label{leq-oim-proc}{\bf (Coherence of $\leq$ and process)}
Let $\pi = (C, \rho)$ be a process of $N(k_0)$ and, moreover, let
$(k_0, \leq_0) \llbracket \sigma, \delta |\rangle (k, \leq)$ be a process sequence for $\pi$.
Then, for all $b, b' \in Max(C)$ we have: 

\[   
\delta(b) \leq \delta(b') \; \iff
     \begin{cases}
        &b \in Min(C) \qquad\qquad\qquad\qquad\;\;\;\;\text{(1)}\\
        &\text{or} \\
        &\pre{b} \neq \emptyset \wedge \pre{b'} \neq \emptyset \wedge \pre{b} \leq_{\pi} \pre{b'} 
        \qquad\;\text{(2)}
     \end{cases}
\]
\proof
We prove the implication in the two directions.
First, we prove that the antecedent implies the consequent by induction on the length of $\sigma$.
    \begin{itemize}
        \item case 0: $\sigma = \epsilon$. \\
            Since $init(N) \llbracket \epsilon, \delta_0 |\rangle init(N)$, we have $C = C^0$ and $b \in Max(C^0) = Min(C^0)$.
            Condition (1) is satisfied.
 
         \item case n+1: $\sigma = \lambda e$ where $e \not \in \lambda$. \\
            The induction hypothesis is that 
            $init(N) \llbracket \lambda, \delta |\rangle (k, \leq)$ is a process sequence for $\pi$,
            where the thesis holds for $\oim{}$. 
           The step is $\oim{} \llbracket t \rangle \oimp{}$ and           
            $\pi = (C, \rho) \deriv{e} (C', \rho') = \pi'$ with $\rho'(e) = t$.
            Let $\kboilerplate{\rho'(e)}$ and $\delta'$ defined as in Definition \ref{proc-seq-def}. 
            Then,
             $init(N) \llbracket \lambda e, \delta' |\rangle (k', \leq')$ is a process sequence for $\pi'$. We have to prove the thesis for
             $(k', \leq')$.\\
            The proof is by cases on the definition of 
            $\delta'(b) \leq' \delta'(b')$. 
            We omit trivial cases.
            
            \begin{itemize}
                \item[-] if $\delta'(b) \in k'' \; , \; \delta'(b') \in k' \setminus k''$ 
                    and $\exists$ $b''$ such that $\delta(b'') \in k \setminus k''$
                    where $\delta'(b) = \delta(b) \leq \delta(b'')$:
                    \begin{itemize}
                        \item[+] if $\delta'(b)$ is such that $b \in Min(C')$: \\
                            condition (1) is satisfied.
                            
                        \item[+] if $\delta'(b)$ is such that $b \not\in Min(C')$: \\
                            then $b \not\in Min(C)$, too, so that, since $\delta(b) \leq \delta(b'')$, by induction we have 
                            $\pre{b} \leq_\pi \pre{b''}$.
                            As $b'' \in \pre{e}$
                            and $b' \in \post{e}$, it is true that
                            $\pre{b''} \leq_{\pi'} \pre{b'}$, and 
                            by transitivity $\pre{b} \leq_{\pi'} \pre{b'}$.
                            Therefore, condition (2) is satisfied.
                    \end{itemize}
            \end{itemize}
    \end{itemize}

    Then, we prove that the consequent implies the antecedent by induction on the length of $\sigma$.
    \begin{itemize}
        \item case 0: $\sigma = \epsilon$. \\
            Since each $b \in Max(C^0)$ is minimal, 
          we have that $\delta(b) \leq p$ for all $p \in k_0$ because $\leq = k_0 \times k_0$.
            
        \item case n+1: $\sigma = \lambda e$ where $e \not \in \lambda$. \\
           The induction hypothesis is that $init(N) \llbracket \lambda, \delta |\rangle (k, \leq)$ is a process sequence for $\pi$, 
           where the thesis holds for $\oim{}$ and $\pi$. The step is $\oim{} \llbracket t \rangle \oimp{}$ and           
            $\pi = (C, \rho) \deriv{e} (C', \rho') = \pi'$ with $\rho'(e) = t$.
            Let $\kboilerplate{\rho'(e)}$ and $\delta'$ defined as in Definition \ref{proc-seq-def}. 
            Then,
             $init(N) \llbracket \lambda e, \delta' |\rangle (k', \leq')$ is a process sequence for $\pi'$. We have to prove the thesis for
             $(k', \leq')$ and $\pi'$. The proof is by inspection on the hypotheses:
            
            \begin{itemize}
                \item[-] if condition (1) holds:
                    since $\delta'(b)$ is minimal for $\leq'$ by Lemma \ref{leq-minimal}, the thesis follows.
                    
                \item[-] if condition (2) holds:
                    there are 4 possible combinations of $\delta'(b), \delta'(b')$. We omit trivial cases.
                    \begin{itemize}
                        \item[+] if $\delta'(b) \in k''$ and $\delta'(b') \in k' \setminus k''$: \\
                            Then, since $\delta'(b') \in k' \setminus k''$, 
                            by Proposition \ref{preset-not-eq-pi}, 
                            it is true that there exists 
                            $b'' \in \pre{e}$ such that
                            $\pre{b} \leq_{\pi} \pre{b''}$,
                            and 
                            $\delta(b'') \in k \setminus k''$.
                            Then, by inductive hypothesis,
                            $\delta'(b) = \delta(b) \leq \delta(b'')$. 
                            Therefore, since $\delta'(b') \in k' \setminus k''$ and $\delta(b'') \in k \setminus k''$, 
                            we have
                            $\delta'(b) \leq' \delta'(b')$.
                            
                        \item[+] if $\delta'(b') \in k''$ and $\delta'(b) \in k' \setminus k''$: 
                            absurd, since $\pre{b} \leq_{\pi'} \pre{b'}$. \qedhere
                    \end{itemize}
            \end{itemize} 
    \end{itemize}
\end{thm}

Note that this theorem holds even in case $\pi = (C, \rho)$ is a {\em partial} process of $N(k_0)$, with a proof that
is a minor adaptation of the one given above.

\begin{thm} \label{c-leq-proc-oim-3-4}
Let $\pi = (C, \rho)$ be a (partial) process of $N(k_0)$ such that $init(N) \llbracket \sigma, \delta |\rangle (k, \leq)$ is a 
process sequence for $\pi$. 
We have that $(k, \leq)$ moves to $(k', \leq')$ through the transition $t$ if and only if 
$\pi$ moves to $\pi'$ through some event $e$, which is mapped to $t$, and
$\sigma e$ is a process sequence for $\pi'$.
More formally, $\oim{} \imtrans{t} \oimp{}$ if and only if, for some $e$, we have ($\pi \derivp{e} \pi'$ or)
$\pi \deriv{e} \pi'$, where  $\rho'(e) = t$ and
$init(N) \llbracket \sigma e, \delta' |\rangle (k', \leq')$ is a process sequence for $\pi'$.

\proof 
By hypothesis, $init(N) \llbracket \sigma, \delta |\rangle (k, \leq)$ is a 
process sequence for $\pi = (C, \rho)$. Then, we prove the two implications separately.

$\Rightarrow$) 
If $\oim{} \imtrans{t} \oimp{}$, then we can extend $\pi$ to $\pi'$ through some event $e$, which is mapped to $t$:
$\pi \deriv{e} \pi'$ (or $\pi \derivp{e} \pi'$), with  $\rho'(e) = t$ and $k''  = k \setminus \delta(\pre{e})$ and  $k' = k'' \boxplus \post{\rho'(e)}$.
By Definition \ref{token-game-oim}, we have that $\oimp{}$ is computed exactly as required
by the definition of process sequence for $\pi'$; indeed,
$init(N) \llbracket \sigma e, \delta' |\rangle (k', \leq')$ is a process sequence for $\pi'$, with $\delta'$ computed as in Definition \ref{proc-seq}.

$\Leftarrow$) 
If $\pi \deriv{e} \pi'$ (or $\pi \derivp{e} \pi'$), with  $\rho'(e) = t$
and $init(N) \llbracket \sigma e, \delta' |\rangle (k', \leq')$ is a process sequence for $\pi'$, then
we observe that $k'' = k \setminus \delta(\pre{e})$ and $k' = k'' \boxplus \post{\rho'(e)}$. This implies
that $\oimp{}$ is computed, according to Definition \ref{proc-seq}, exactly as required by 
Definition \ref{token-game-oim}, so that $\oim{} \imtrans{t} \oimp{}$.
\qed
\end{thm}

\begin{exa}\label{leq_and_leq_pi_bounded_example}
In Figure \ref{fig:leq_and_leq_pi_bounded_example}(a), 
the same 5-bounded P/T net $N$ as Figure \ref{fig:token-game-1} is depicted, together with its empty process (we omit to represent its initial marking). 
Figure \ref{fig:leq_and_leq_pi_bounded_example}(b,c) shows how the process corresponding to the transition sequence $t_2 \, t_1$ grows. 
We consider the same execution as in Example \ref{token-game-im-example}, i.e. 
$k_0 \imtrans{t_2} k_1 \imtrans{t_1} k_2$.
For simplicity's sake, in the following 
each condition will be mapped to the place having same subscript and each
event will be mapped to the transition having same label.
We will denote each process $\pi_i$ as the one thus corresponding to causal net $C_i$.
Before any transition fires, we have $init(N) = (k_0, \leq_0)$ where 
$\leq_0 = k_0 \times k_0$ by Definition \ref{oim}.

Not surprisingly, all conditions $b_i^j$ are minimal in the causal net $C^0$ and 
mapped to tokens in $(k_0, \leq_0)$. 
The firing of $t_2$ deletes token $(s_2, 2)$ and generates token $(s_3, 1)$;
moreover, since $(s_2, 1) \leq_0 (s_2, 2)$ we have $(s_2, 1) \leq_1 (s_3, 1)$.
Note that $b_2^1 \in Min(C_1)$ but $b_3 \not\in Min(C_1)$.
After the firing of $t_1$, there are four tokens in place $s_2$. 
However, since $(s_2,2)$ and $(s_2, 4)$ are generated by $t_1$,
they are greater in $\leq_2$ than $(s_2, 1)$ and $(s_2, 3)$.
This can also be seen at the process level: $b_2^1$ and $b_2^3$ are minimal conditions of $C_2$,
while $b_2^4$ and $b_2^5$ are not.
On the other hand, note that, just as 
$b_2^4$ and $b_3$ are not minimal in $C_2$ but also not related by $\leq_{\pi_2}$,
also $(s_2, 2)$ and $(s_3, 1)$ are not related by $\leq_2$.
\end{exa}

\section{Fully-concurrent Bisimilarity is Decidable} \label{fc-dec-sec}

We now define a novel bisimulation relation based on ordered indexed markings (oim, for short), 
generalizing the similar idea in \cite{Vog91}. 

An OIM bisimulation is a relation composed of triples of the form $\oimt$, such that relation $\beta \subseteq k_1 \times k_2$ relates
tokens of the two indexed markings $k_1$ and $k_2$. The initial triple of an OIM bisimulation is $(init(N(k_1)), init(N(k_2)), k_1 \times k_2)$.
Then, whenever the first oim moves with a transition $t_1$, the second oim must respond with a transition $t_2$ such that not only
the label of the two transitions is the same, but also 
the two transitions must consume individual tokens related via $\beta$.
As individual tokens only interest us as far as precedences in their generation are
concerned, we do not require that the tokens consumed by $t_1$ are in a bijective
correspondence to those consumed by $t_2$; we do not even require that an individual token
consumed by the first transition is itself related to another consumed by the second one;
it is enough that each token consumed by $t_1$ precedes
a (possibly different) token consumed by $t_1$
that is related via $\beta$ to a token consumed by $t_2$: this allows preserving causality among the 
generated events.

Moreover, if $\oimmvs{1}{t_1}$ and $(k_2, \leq_2)$ responds with $\oimmvs{2}{t_2}$, then it is required 
that $\oimtp$ is in the OIM bisimulation,
where the new relation $\beta'$ is obtained from $\beta$ by retaining all the pairs of individual 
tokens related by $\beta$ but untouched by the two transitions, and by adding
all the pairs of individual tokens generated by the two matching transitions.

\begin{defi}\label{oim-bis}{\bf (OIM bisimulation) }
Let $N = (S, A, T)$ be a P/T net. An OIM bisimulation is a relation 
$\mathfrak{B} \subseteq OIM(N) \times OIM(N) \times \mathcal{P}( (S \times \nat) \times (S \times \nat))$
such that if  $\oimt \in \mathfrak{B}$, then:

\begin{figure}[ht]
    \centering
    
    \begin{tikzpicture}[
        every place/.style={draw,thick,inner sep=0pt,minimum size=6mm},
        every transition/.style={draw,thick,inner sep=0pt,minimum size=4mm},
        bend angle=30,
        pre/.style={<-,shorten <=1pt,>=stealth,semithick},
        post/.style={->,shorten >=1pt,>=stealth,semithick}    
    ]
    
    \node (a) [label=left:$a)$]{};

    \node (p1) [place]  [right = {1cm} of a, label=above:$s_1$] {}
        [children are tokens]
            child {node [token] {1}};
    \node (t1) [transition] [below of = p1,label=left:$u$] {};
    \node (p2) [place] [below of = t1,label=left:$s_2$] {}
        [children are tokens]
            child {node [token] {1}}
            child {node [token, fill=red] {2}}
            child {node [token] {3}};
    
    \node (t2) [transition] [below of = p2,label=left:$v$] {};
    \node (p3) [place, tokens=0] [below of = t2,label=left:$s_3$] {};

    \draw  [->] (p1) to (t1);
    \draw  [->] (t1) to node[auto,swap] {2} (p2);
    \draw  [->] (p2) to (t2);
    \draw  [->] (t2) to (p3);
    
    \node (t) [below = {0.9cm} of p3, label={\begin{rotate}{-90} $\imtrans{t_2}$ \end{rotate}}] {};

    \node (ap) [right={10cm} of p1, label=left:$(C^0$]{};
    \node (b1) [place]  [left={4cm} of ap, label=above:$b_1$] {};   
    \node (b2) [place]  [right of = b1, label=above:$b_2^1$] {};
    \node (b3) [place]  [right of  = b2, label=above:$b_2^2$] {};
    \node (b4) [place]  [right of = b3, label=above:$b_2^3$] {};

    \node (et) [above right = {1cm} and {7.8cm} of t,label={\begin{rotate}{-90} $\trns{e_v}$ \end{rotate}}] {};

    \draw  [dashed, ->, bend right] (b1) to (p1);
    \draw  [dashed, ->, bend left] (b2) to (p2);
    \draw  [dashed, ->, bend left] (b3) to (p2);
    \draw  [dashed, ->, bend left] (b4) to (p2);
    \end{tikzpicture}

    \begin{tikzpicture}[
        every place/.style={draw,thick,inner sep=0pt,minimum size=6mm},
        every transition/.style={draw,thick,inner sep=0pt,minimum size=4mm},
        bend angle=30,
        pre/.style={<-,shorten <=1pt,>=stealth,semithick},
        post/.style={->,shorten >=1pt,>=stealth,semithick}    
    ]
    
    \node (a) [label=left:$b)$]{};

    \node (p1) [place]  [right = {1cm} of a, label=above:$s_1$] {}
        [children are tokens]
            child {node [token, fill=red] {1}};
    \node (t1) [transition] [below of = p1,label=left:$u$] {};
    \node (p2) [place] [below of = t1,label=left:$s_2$] {}
        [children are tokens]
            child {node [token] {1}}
            child {node [token] {3}};
    \node (t2) [transition] [below of = p2,label=left:$v$] {};
    \node (p3) [place, tokens=1] [below of = t2,label=left:$s_3$] {}
        [children are tokens]
            child {node [token, fill = blue] {1}};

    \draw  [->] (p1) to (t1);
    \draw  [->] (t1) to node[auto,swap] {2} (p2);
    \draw  [->] (p2) to (t2);
    \draw  [->] (t2) to (p3);
    
    \node (t) [below = {0.9cm} of p3, label={\begin{rotate}{-90} $\imtrans{t_1}$ \end{rotate}}] {};

    \node (ap) [right={10cm} of p1, label=left:$(C_1$]{};
    \node (b1) [place]  [left={4cm} of ap, label=above:$b_1$] {};   
    \node (b2) [place]  [right of = b1, label=above:$b_2^1$] {};
    \node (b3) [place]  [right of  = b2, label=above:$b_2^2$] {};
    \node (b4) [place]  [right of = b3, label=above:$b_2^3$] {};
    \node (e1) [transition] [below of = b3, label=right:$e_v$] {};
    \node (b5) [place]  [below of = e1, label=left:$b_3$] {};

    \draw  [->] (b3) to (e1);
    \draw  [->] (e1) to (b5);

    \node (et) [above right = {1cm} and {7.8cm} of t,label={\begin{rotate}{-90} $\trns{e_u}$ \end{rotate}}] {};

    \draw  [dashed, ->, bend right] (b1) to (p1);
    \draw  [dashed, ->, bend left] (b2) to (p2);
    \draw  [dashed, ->, bend left] (b5) to (p3);
    \draw  [dashed, ->, bend left] (b4) to (p2);
    \end{tikzpicture}

    \begin{tikzpicture}[
        every place/.style={draw,thick,inner sep=0pt,minimum size=6mm},
        every transition/.style={draw,thick,inner sep=0pt,minimum size=4mm},
        bend angle=30,
        pre/.style={<-,shorten <=1pt,>=stealth,semithick},
        post/.style={->,shorten >=1pt,>=stealth,semithick}    
    ]
    
    \node (a) [label=left:$c)$]{};

    \node (p1) [place,tokens=0]  [right = {1cm} of a, label=above:$s_1$] {};
    \node (t1) [transition] [below of = p1,label=left:$u$] {};
    \node (p2) [place] [below of = t1,label=left:$s_2$] {}
        [children are tokens]
        child {node [token] {1}}
        child {node [token, fill = blue] {2}}
        child {node [token] {3}}
        child {node [token, fill = blue] {4}};

    \node (t2) [transition] [below of = p2,label=left:$v$] {};
    \node (p3) [place, tokens=1] [below of = t2,label=left:$s_3$] {}
        [children are tokens]
        child {node [token] {1}};

    \draw  [->] (p1) to (t1);
    \draw  [->] (t1) to node[auto,swap] {2} (p2);
    \draw  [->] (p2) to (t2);
    \draw  [->] (t2) to (p3);

    \node (ap) [right={10cm} of p1, label=left:$(C_2$]{};
    \node (b1) [place]  [left={4cm} of ap, label=above:$b_1$] {};   
    \node (b2) [place]  [right of = b1, label=above:$b_2^1$] {};
    \node (b3) [place]  [right of  = b2, label=above:$b_2^2$] {};
    \node (b4) [place]  [right of = b3, label=above:$b_2^3$] {};
    \node (e1) [transition] [below of = b3, label=right:$e_v$] {};
    \node (b5) [place]  [below of = e1, label=below:$b_3$] {};
    \node (e2) [transition] [below of = b1, label=right:$e_u$] {};
    \node (b6) [place]  [below left of = e2, label=below:$b_2^4$] {};
    \node (b7) [place]  [below right of = e2, label=below:$b_2^5$] {};

    \draw  [->] (b3) to (e1);
    \draw  [->] (e1) to (b5);
    \draw  [->] (b1) to (e2);
    \draw  [->] (e2) to (b6);
    \draw  [->] (e2) to (b7);

    \draw  [dashed, ->] (b6) to (p2);
    \draw  [dashed, ->, bend left] (b7) to (p2);
    \draw  [dashed, ->, bend right] (b2) to (p2);
    \draw  [dashed, ->, bend left] (b5) to (p3);
    \draw  [dashed, ->] (b4) to (p2);
    \end{tikzpicture}

    \caption{Execution of the transition labeled by $v$, then $u$, on the net of Figure \ref{fig:token-game-1} and corresponding process (only the mapping of maximal conditions to tokens is displayed). Tokens to be consumed are red, generated ones blue.}
    \label{fig:leq_and_leq_pi_bounded_example}
\end{figure}

\clearpage 

\begin{itemize}
    \item $\forall t_1, k_1', \leq_1'$ such that $\oimmvs{1}{t_1}$, 
    (where we assume $\kboilerplateindex{t_1}{1}$),
    there exist $t_2, k_2', \leq_2'$
    (where we assume $\kboilerplateindex{t_2}{2}$),
    and for $\beta'$ defined as:
    $ \forall p_{1} \in k_1', \forall p_{2} \in k_2'$
    \[   
        p_{1} \; \beta' \; p_{2} \iff 
         \begin{cases}
            &p_{1} \in \old{1} \, , \, p_{2} \in \old{2} \text{ and }
            p_{1} \mathrel{\beta} p_{2} \\
            &\text{or} \\
            &p_{1} \in \generated{1} \, , \, p_{2} \in \generated{2}
         \end{cases}
    \]   
        the following hold:
        \begin{itemize}
            \item[-] $\oimmvs{2}{t_2}$ where $\oimtp \in \mathfrak{B}$ and $l(t_1) = l(t_2)$;
            \item[-] $\forall p_{1} \in \deleted{1}$, 
            $\exists p_{1}' \in \deleted{1} \, , \, \exists p_{2}' \in \deleted{2}$ such that
            $p_{1} \leq_1 p_{1}' \, \wedge \, p_{1}' \mathrel{\beta} p_{2}'$ 
            and, symmetrically, 
            $\forall p_{2} \in \deleted{2}$, 
            $\exists p_{2}' \in \deleted{2} \, , \, \exists p_{1}' \in \deleted{1}$ such that
            $p_{2} \leq_2 p_{2}' \, \wedge \, p_{1}' \mathrel{\beta} p_{2}'$ 
        \end{itemize}
        
    \item symmetrically, if $\oim{2}$ moves first.
\end{itemize}

Two markings $m_1$ and $m_2$ of $N$ are OIM bisimilar, 
denoted $m_1 \sim_{oim} m_2$, 
if there exists an OIM bisimulation $\mathfrak{B}$ 
containing the triple 
$(init(N(k_1)), init(N(k_2)), k_1 \times k_2)$ 
where, for $i = 1, 2$, $k_i$ is the initial (i.e., closed) indexed marking such that $m_i = \alpha(k_i)$.
\end{defi}

Next, we show that fully-concurrent bisimilarity $\sim_{fc}$ and OIM-bisimilarity $\sim_{oim}$ coincide on P/T nets, by first
proving that fully-concurrent bisimilarity implies OIM-bisimilarity,
 and then by proving that OIM-bisimilarity implies fully-concurrent bisimilarity.
 The basic idea behind these proofs is that two tokens are related by $\beta$
if and only if the transition generating one of the two is mapped by the order-isomorphism $f$ to the transition generating the other one.

\begin{thm}\label{fc-implies-oim-bounded} {\bf (FC-bisimilarity implies OIM-bisimilarity)}
Let $N = (S,A,T)$ be a net.
Given two markings $m_{01}, m_{02}$ of $N$,
if $m_{01} \sim_{fc} m_{02}$, then $m_{01} \sim_{oim} m_{02}$. 

\proof

If $m_{01} \; \sim_{fc} \; m_{02}$, then there exists an fc-bisimulation $R_1$ 
containing the triple $(\pi^0_1, \emptyset, \pi^0_2)$, 
where $\pi^0_i = (C^0_i, \rho^0_i)$ is such that $C^0_i$ contains no events and 
$\rho^0_i(Min(C^0_i))  = \rho^0_i(Max(C^0_i)) = m_{0i}\;$ for $i = 1, 2$.

Given closed indexed markings $k_{0i}$ such that $m_{0i} = \alpha(k_{0i})$
for $i = 1, 2$,
let us consider 
    \begin{equation*} \label{R2-OIM-BOUNDED}
        \begin{split}
        R_2 \overset{def}{=} \{ \oimt | & \fct \in R_{1} \text{ and} \\
         & \text{for } i = 1,2 \text{, } \pi_i \text{ is a process of } N(k_{0i}) \text{ and} \\
         & init(N(k_{0i})) \llbracket \sigma_i, \delta_i |\rangle \oim{i} \text{ is a process sequence for } \pi_i \text{ and}\\
         &f(\sigma_1) = \sigma_2 \text{ and}\\
         & \forall p_1 \in k_1, \forall p_2 \in k_2: 
          p_{1} \mathrel{\beta} p_{2} \text{ if and only if } \\
         &\qquad \exists b_1 \in Max(C_1) \text{ such that }\delta_1(b_1) = p_1 \text{ and}\\
         &\qquad \exists b_2 \in Max(C_2) \text{ such that }\delta_2(b_2) = p_2 \text{ and}\\
         & \qquad \text{either } b_1 \in Min(C_1) \wedge b_2 \in Min(C_2) \\
         & \qquad \text{or } \pre{b_1} \neq \emptyset \wedge
         \pre{b_2} \neq \emptyset \wedge 
         f( \pre{b_1}) = \pre{b_2}
         \}.
        \end{split}
    \end{equation*}
If we prove that $R_2$ is an OIM bisimulation, then since $(\pi^0_1, \emptyset, \pi^0_2) \in R_1$, 
$init(N(k_{0i})) \llbracket \epsilon, \delta_{0i} |\rangle $ \\ $ init(N(k_{0i}))$ is a process sequence for $\pi^0_i$ and $\rho^0_i(Min(C^0_i)) 
= \rho^0_i(Max(C^0_i)) = k_{0i}$ where $\alpha(k_{0i}) = m_{0i}$ for $i = 1, 2$, by definition of $R_2$ it follows that 
$(init(N(k_{01})), init(N(k_{02})), k_{01} \times k_{02}) \in R_2$ and therefore $m_{01} \sim_{oim} \; m_{02}$.

Assume $\oimt \in R_2$. By symmetry, we consider only the case when $(k_1, \leq_1)$ moves first.
Let $\oimmvs{1}{t_1}$.
By definition of $R_2$, there is a process sequence $init(N(k_{01})) \llbracket \sigma_1, \delta_1 |\rangle \oim{1}$ 
for $\pi_1$. Hence, by Theorem \ref{c-leq-proc-oim-3-4}, it follows that
$\pi_1 \deriv{e_1} \pi_1'$ with $\rho_1'(e_1) = t_1$, and that 
$init(N(k_{01})) \llbracket \sigma_1 e_1, \delta_1' |\rangle \oimp{1}$ is a process sequence for $\pi_1'$. 

As $\fct \in R_1$, it follows that $\pi_2 \deriv{e_2} \pi_2'$ with $\rho_2'(e_2) = t_2$ 
and $\fctp \in R_1$, where $f'$ extends $f$ by $f'(e_1) = e_2$. 
By def. of $R_2$, a process sequence
$init(N(k_{02})) \llbracket \sigma_2, \delta_2 |\rangle \oim{2}$ for $\pi_2$ exists; moreover, 
$init(N(k_{02})) \llbracket \sigma_2 e_2, \delta_2' |\rangle \oimp{2}$ is a process sequence for $\pi_2'$, 
so that, by Theorem \ref{c-leq-proc-oim-3-4}, we have $\oimmvs{2}{t_2}$.

Summing up, we have that $\fctp \in R_1$, $init(N(k_{01})) \llbracket \sigma_1 e_1, \delta_1' |\rangle \oimp{1}$ is a 
process sequence for $\pi_1'$,
$init(N(k_{02})) \llbracket \sigma_2 e_2, \delta_2' |\rangle \oimp{2}$ is a process sequence for $\pi_2'$,
$f'(\sigma_1 e_1) = \sigma_2 e_2$, and moreover, for $\beta'$ defined as follows:
    \begin{equation*} 
        \begin{split}
        & \forall p_1 \in k'_1, \forall p_2 \in k'_2: 
          p_{1} \mathrel{\beta}' p_{2} \text{ if and only if } \\
         &\qquad \exists b_1 \in Max(C'_1) \text{ such that }\delta_1'(b_1) = p_1 \text{ and},\\
         &\qquad \exists b_2 \in Max(C'_2) \text{ such that }\delta_2'(b_2) = p_2 \text{ and},\\
         & \qquad \text{either } b_1 \in Min(C'_1) \wedge b_2 \in Min(C'_2) \\
         & \qquad \text{or } \pre{b_1} \neq \emptyset \wedge
         \pre{b_2} \neq \emptyset \wedge 
         f'( \pre{b_1}) = \pre{b_2}
         \},
        \end{split}
    \end{equation*}
we get that $\oimtp \in R_2$ by definition of $R_2$.

Therefore, we have proved that to the move $\oimmvs{1}{t_1}$, $(k_2, \leq_2)$ can reply with the move
$\oimmvs{2}{t_2}$, where $\oimtp \in R_2$ and $l(t_1) = l(t_2)$.
Hence, in order to prove that $\oimt$ is a OIM bisimulation triple, as required, 
it remains to prove that 
the definition of $\beta'$ arising from $R_2$, i.e., the unique $\beta'$ such that the triple $((k_1', \leq_1'), (k_2', \leq_2'), \beta') \in R_2$
for $(\pi_1', f', \pi_2') \in R_1$, 
is coherent with the one of Definition \ref{oim-bis}, 
i.e., it implies both
    \begin{enumerate}
        \item $\forall p_{1} \in k_1' \, , \, p_{2} \in k_2' \, $
    \[   
        p_{1} \mathrel{\beta'}  p_{2} \iff
         \begin{cases}
            &p_{1} \in \old{1} \, , \, p_{2} \in \old{2} \text{ and }
            p_{1} \mathrel{\beta} p_{2} \quad\qquad \text{(i)}\\
            &\text{or} \\
            &p_{1} \in \generated{1} \, , \, p_{2} \in \generated{2} 
            \qquad\qquad\;\;\text{(ii)}
         \end{cases}
    \]
 and
        \item $\forall p_{1} \in \deleted{1}$, 
            $\exists p_{1}' \in \deleted{1},  \, \exists p_{2}' \in \deleted{2}$ such that
            $p_{1} \leq_1 p_{1}' \, \wedge \, p_{1}' \mathrel{\beta} p_{2}'$, 
            and symmetrically 
            $\forall p_{2} \in \deleted{2}$, 
            $\exists p_{2}' \in \deleted{2},  \, \exists p_{1}' \in \deleted{1}$ such that
            $p_{2} \leq_2 p_{2}' \, \wedge \, p_{1}' \mathrel{\beta} p_{2}'$ . 
    \end{enumerate}

    \begin{itemize}
        \item[] {\it Proof 1)}
            The two implications are proved separately.
            \begin{itemize}
                \item[] {\it Proof $\xRightarrow{}$}: 
                assume $p_1 = \delta_1'(b_1)$ and $p_2 = \delta_2'(b_2)$. Then:
                \begin{itemize}
                    \item[-] If $b_1 \in Min(C_1')$ and $b_2 \in Min(C_2')$: \\
                        then for $i=1,2$,  $b_i \in Max(C_i)$, too. This implies
                        $p_{i} \in \old{i}$ and $p_{1} \beta p_{2}$, 
                        satisfying condition (i).
                     
                    \item[-] if $\pre{b_1} \neq \emptyset \wedge
                                \pre{b_2} \neq \emptyset \wedge 
                                f'(\pre{b_1}) = \pre{b_2}$: \\
                                Let us consider events $e_1', e_2'$
                                such that $b_1 \in \post{e_1'}$ and 
                                $b_2 \in \post{e_2'}$.
                                There are four possible cases: 
                                \begin{itemize}
                                    \item[+] if $e_1' = e_1$ and $e_2' = e_2$:
                                      since $p_1 = \delta_1'(b_1)$ and 
                                      $t_1 = \rho_1'(e_1) = \rho_1'(e_1')$, 
                                      we have $p_{1} \in \generated{1}$.
                                      For the same reason, $p_{2} \in \generated{2}$ and
                                      therefore condition (ii) holds.
                                  
                                    \item[+] if $e_1' \neq e_1$ and $e_2' \neq e_2$:
                                    then $b_1$ is maximal also in $C_1$ because $e_1'$ occurred before $e_1$, hence
                                    $p_{1} \in \old{1}$.
                                        For the same reason, 
                                        $p_{2} \in \old{2}$.
                                        Since $\oimct \in R_2$ and
                                        $f(\pre{b_1}) = \pre{b_2}$, 
                                        then $p_{1} \beta p_{2}$, 
                                        therefore condition (i) holds.
                                   
                                    \item[+] other cases: absurd
                                    since $f'(e_1) = e_2$.
    
                                \end{itemize}
                \end{itemize}
  
                  \item[] {\it Proof $\xLeftarrow{}$}: 
                Since $p_{1} \in k_1'$ and 
                $p_{2} \in k_2'$, there exist
                $b_1$ and $b_2$ such that
                $p_1 = \delta_1'(b_1)$ and $p_2 = \delta_2'(b_2)$. 
                
               \begin{itemize}
                            \item[-] if $p_{1} \in \old{1} \, , \, 
                            p_{2} \in \old{2} \text{ and }
                            p_{1} \mathrel{\beta} p_{2}$ : 
                            
                            We need to separate two cases for $p_1$.
                            \begin{enumerate}
                                \item[+] if $b_{1} \in Min(C_1')$: 
                                    since $p_{1} \mathrel{\beta} p_{2}$, 
                                    it is true that $b_2 \in Min(C_2')$.
                                
                                \item[+] if $b_{1} \not\in Min(C_1')$: 
                                    then $\pre{b_1} \neq \emptyset$ and, as  $p_{1} \mathrel{\beta} p_{2}$, also
                                    $\pre{b_2} \neq \emptyset$. 
                                    As
                                    $p_{1} \in \old{1}$,
                                    $p_{2} \in \old{2}$, and 
                                    $\oimct \in R_2$, 
                                    we have $b_i \in Max(C_i)$ and  
                                    $f(\pre{b_1)} = \pre{b_2}$, 
                                    and by conservative extension of $f$ to $f'$ we get the thesis.
                            \end{enumerate}
                            
                            \item[-] if $p_{1} \in \generated{1} \, , \, p_{2} \in \generated{2}$ : \\
                            then since $\delta'_1(b_1) = p_1$ we have $\pre{b_1} = e_1$, 
                            therefore $\pre{b_1} \neq \emptyset$. 
                            The same applies to $p_2$, and since 
                            $f'(e_1) = e_2$, we have $f'(\pre{b_1}) = \pre{b_2}$.
                        
                        \end{itemize}
             \end{itemize}
                     
         \item[] {\it Proof 2)}
        Let us consider events $e_1, e_2$ such that 
        $\pi_1 \deriv{e_1} \pi'_1$, 
        $\pi_2 \deriv{e_2} \pi'_2$ 
        and
        $f'(e_1) = e_2$, with $(\pi_1', f', \pi_2') \in R_1$.
        We assume a token $p_{1} \in \deleted{1}$ such that 
        there exists $b_1$ with $\delta_1(b_1) = p_1$.
        Note that, since $\pi_1 \deriv{e_1} \pi'_1$,  
        it is true that $b_1 \in Max(C_1)$ and $b_1 \in \pre{e_1}$.
        We are to prove that $\exists p'_{1} \in \deleted{1} \, , \,
        \exists p'_{2} \in \deleted{2}$ such that
        $p_{1} \leq_1 p'_{1}$ and $p'_{1} \mathrel{\beta} p'_{2}$.
        In the following, in some cases we have
        $p_{1} = p'_{1}$: if that is true, then 
        $p_{1} \leq_1 p'_{1}$ by reflexivity of $\leq_1$.

            There are two possible cases for $b_1$:
            \begin{itemize}
                \item[-] if  $b_1 \in Min(C_1)$:\\ 
                    There are two possible subcases:
                    \begin{itemize}
                        \item[+] $\exists b_2' \in \pre{e_2}$ such that 
                        $b_2' \in Min(C_2)$: \\
                        Then by definition of $\delta_2$ there exists a token $p_2' = \delta_2(b_2')$.
                        By definition of $\beta$, we have
                        $p_{1} \mathrel{\beta} p'_{2}$.
                        
                        \item[+] otherwise: \\
                             Since  $\pi_2 \deriv{e_2} \pi'_2$, 
                            there exists a condition $b_2' \in Max(C_2)$ such that 
                            $\delta_2(b_2') = p'_2$, where token $p'_{2}$  $\in \deleted{2}$. 
                           Let us consider 
                            the event $e_2' \in \mathsf{E}_{C_2}$ such that $e_2' = \pre{b_2'}$.                             
                            Since $f$ is an isomorphism between 
                            $\mathsf{E}_{C_1}$ and $\mathsf{E}_{C_2}$, there exists
                            event $e_1' \in \mathsf{E}_{C_1}$ such that $f(e_1') = e_2'$; moreover,
                            there exists $b_1' \in Max(C_1)$ such that $e_1' = \pre{b_1'}$ and $\delta_1(b_1') = p_1'$.
                            Since $b_1 \in Min(C_1)$, by Lemma \ref{leq-minimal}
                            it is true that $p_{1}$ is minimal for $\leq_1$, and
                            therefore $p_{1} \leq_1 p'_{1}$.
                            Finally, since $e_1' = \pre{b_1'}$,
                            $e_2' = \pre{b_2'}$ and
                            $f(e_1') = e_2'$,
                            we have $p'_{1} \mathrel{\beta} p'_{2}$.
                    \end{itemize}
                    
                  \item[-] if $b_1 \not\in Min(C_1)$: \\
                    Let $p_1' = \delta_1(b_1') \in k_1 \setminus k_1''$ such that $\pre{b_1} \leq_{\pi_1} \pre{b_1'}$
                    and  $\not \exists p = \delta_1(b) \in k_1 \setminus k_1''$ such that $\pre{b_1'} <_{\pi_1} \pre{b}$.
                    Hence, $p_1 \leq_1 p_1'$ and $\pre{b_1'} = e_1'$ is a maximal proper predecessor of $e_1$ in $\mathsf{E}_{C'_1}$.
                    Since $f$ is an isomorphism between
                    $\mathsf{E}_{C_1}$ and $\mathsf{E}_{C_2}$,
                    there exists $e_2' \in \mathsf{E}_{C_2}$
                    such that $e_2' = f(e_1')$.
                    Since $e_1'$ is an immediate predecessor of $e_1$
                    in $\mathsf{E}_{C_1'}$, by definition of isomorphism $f'$, 
                    it is true that also $e_2'$ is an immediate
                    predecessor of $e_2$ in $\mathsf{E}_{C_2'}$. 
                    Therefore,
                    it is possible to choose a 
                    condition $b_2'$, with $\delta_2(b_2') = p'_{2}$, such that not only
                    $e_2' = \pre{b_2'}$, 
                    but also $b_2' \in \pre{e_2}$.
                     Finally, we have  $e_1' = \pre{b_1'}$,
                    $e_2' = \pre{b_2'}$ and
                    $f(e_1') = e_2'$,
                    hence
                    $p_{1}' \mathrel{\beta} p'_{2}$.
        
            \end{itemize}
            The proof of the case for $p_{2} \in \deleted{2}$ 
            is symmetrical and therefore omitted.                 
        \end{itemize}

As mentioned above, the case in which $(k_2, \leq_2)$ moves first is symmetrical and so omitted. 
Therefore, $R_2$ is an OIM bisimulation and $m_{01} \sim_{oim} m_{02}$.    
\qed
\end{thm}

Now we prove the reverse implication, so that, at the end, we get that $\sim_{oim}$ and $\sim_{fc}$ coincide.
 
\begin{thm}\label{oim-implies-fc-bounded} {\bf (OIM-bisimilarity implies FC-bisimilarity)}
Let $N = (S,A,T)$ be a net. 
Given two markings $m_{01}, m_{02}$ of $N$,
if $m_{01} \; \sim_{oim} \; m_{02}$, then $m_{01} \; \sim_{fc} \; m_{02}$.
\proof
If $m_{01} \; \sim_{oim} \; m_{02}$, then there exists an OIM bisimulation $R_1$ containing the triple 
$(init(N(k_{01})),$ $ init(N(k_{02})), k_{01} \times k_{02})$, 
where $\alpha(k_{01}) = m_{01}$, $\alpha(k_{02}) = m_{02}$, and $k_{01}, k_{02}$ are closed.

\noindent
Let us consider 
    \begin{equation*} \label{R2-fc-from-oim-bounded}
        \begin{split}
        R_2 \overset{def}{=} \{ \fct \mid 
         & \oimct \in R_1 \text{ and} \\
         & \text{for } i = 1,2 \text{, } 
         \pi_i = (C_i, \rho_i) \text{ is a process of } N(k_{0i}) \text{  and} \\
         & f \text{ is an isomorphism $\mathsf{E}_{C_1} \xrightarrow{} \mathsf{E}_{C_2}$ and} \\
         & init(N(k_{0i})) \llbracket \sigma_i, \delta_i |\rangle \oim{i} \text{is a process sequence for } \pi_i \text{ and}\\
         &f(\sigma_1) = \sigma_2 \text{ and}\\
         & \forall p_1 \in k_1, \forall p_2 \in k_2: 
          p_{1} \mathrel{\beta} p_{2} \text{ if and only if } \\
         &\qquad \exists b_1 \in Max(C_1) \text{ such that }\delta_1(b_1) = p_1 \text{ and}\\
         &\qquad \exists b_2 \in Max(C_2) \text{ such that }\delta_2(b_2) = p_2 \text{ and}\\
         & \qquad \text{either } b_1 \in Min(C_1) \wedge b_2 \in Min(C_2) \\
         & \qquad \text{or } \pre{b_1} \neq \emptyset \wedge
         \pre{b_2} \neq \emptyset \wedge 
         f( \pre{b_1}) = \pre{b_2}
         \}.
        \end{split}
    \end{equation*}
If we prove that $R_2$ is an fc-bisimulation, 
then we have that $m_{01} \sim_{fc} m_{02}$. 
In fact, consider for $i = 1, 2$ the empty process $(C^0_i, \rho^0_i)$ 
(i.e., 
$C^0_i$ contains no events and
$\rho^0_i(Min(C^0_i))  = \rho^0_i(Max(C^0_i)) = m_{0i}$).
Since
$(init(N(k_{01})), init(N(k_{02})), k_{01} \times k_{02}) \, \in R_1$ and, for $i = 1, 2$, 
$(C^0_i, \rho^0_i)$ is a process of $N(k_{0i})$ and
$init(N(k_{0i})) \llbracket \epsilon, \delta_{0i} |\rangle init(N(k_{0i}))$ 
is a process sequence for $\pi_{0i}$ and 
$\alpha(k_{0i}) = m_{0i}$, by definition of $R_2$
it follows that $(\pi^0_1, \emptyset, \pi^0_2) \in R_2$, 
and therefore $m_{01} \; \sim_{fc} \; m_{02}$.

Assume $\fct \in R_2$. By symmetry, we consider only the case when $\pi_1$ moves first.
Let $\pi_1 \deriv{e_1} \pi_1'$ 
where $\rho_1'(e_1) = t_1$.

By definition of $R_2$, there is  a process sequence 
$init(N(k_{01})) \llbracket \sigma_1, \delta_{1} |\rangle \oim{1}$ 
for $\pi_1$; moreover, $init(N(k_{01})) \llbracket \sigma_1  e_1 , \delta_{1}' |\rangle \oimp{1}$
is a process sequence for $\pi_1'$. Hence, by Theorem \ref{c-leq-proc-oim-3-4}, 
it follows that $(k_1, \leq_1) \imtrans{t_1} (k_1', \leq_1')$.

As $\oimct \in R_1$, then there exist $t_2, k_2', \leq_2', \beta'$ such that 
$\oimmvs{2}{t_2}$ and 
$\oimctp \in R_1$. 
By definition of $R_2$, a process sequence
$init(N(k_{02})) $\\$\llbracket \sigma_2, \delta_{2} |\rangle \oim{2}$  for $\pi_2$ exists. 
By Theorem \ref{c-leq-proc-oim-3-4} we have 
$\pi_2 \deriv{e_2} \pi_2'$, 
where   $\rho_2'(e_2) = t_2$ and 
$init(N(k_{02})) \llbracket \sigma_2  e_2, \delta_{2}' |\rangle \oimp{2}$ is a process sequence for $\pi_2'$. 
Note that, for $i = 1,2$, $\pi'_i$ is a process of $N(k_{0i})$.

We extend $f$ to $f'$ with the mapping $f'(e_1) = e_2$: 
since $f$ is an isomorphism 
between $\mathsf{E}_{C_1}$ and $\mathsf{E}_{C_2}$, 
and $\pi_1 \deriv{e_1} \pi_1'$, $\pi_2 \deriv{e_2} \pi_2'$,
in order to prove that $f'$ is an isomorphism between $\mathsf{E}_{C_1'} = (E_1', \preceq_1')$ and $\mathsf{E}_{C_2'} = (E_2', \preceq_2')$
such that $f'(\sigma_1 e_1) = \sigma_2 e_2$, we have only to prove that $f$ maps the predecessors of $e_1$ to those of $e_2$ (and viceversa).
For this check we need the following two facts: 
\begin{enumerate}
\item if $e \preceq_1' e_1$ then $f'(e) \preceq_2' f'(e_1)$, where $e$ is an  
event that immediately precedes $e_1$, hence, $e \preceq_1' e_1$ and there exists $b_1$ such that
$b_1 \in \post{e}$ and $b_1 \in \pre{e_1}$;
\item if $e' \preceq_2' e_2$ then $f'^{-1}(e') \preceq_1' f'^{-1}(e_2)$, where $e'$ is an 
event that immediately precedes $e_2$, hence, $e' \preceq_2' e_2$ and there exists $b_2$ such that
$b_2 \in \post{e'}$ and $b_2 \in \pre{e_2}$.
\end{enumerate}
Since these two facts are essentially symmetric, we will prove only the first one.

If $b_1 \in \post{e}$ and $b_1 \in \pre{e_1}$, then $\delta_1(b_1) = p_1 \in k_1 \setminus k_1''$. By Definition \ref{oim-bis},
there exist $p_1' \in k_1 \setminus k_1''$ (hence, $b_1' \in \pre{e_1}$ such that $\delta_1(b_1') = p_1'$)
and $p_2' \in k_2 \setminus k_2''$ (hence, $b_2' \in \pre{e_2}$ such that $\delta_2(b_2') = p_2'$) such that 
$p_1 \leq_1 p_1'$ and $p_1' \beta p_2'$. Since $p_1 \leq_1 p_1'$, we know by Theorem \ref{leq-oim-proc} that 
$\pre{b_1} \leq_{\pi_1} \pre{b_1'}$, i.e., $e \preceq_1  \pre{b_1'}$; this implies that
$f(e) \preceq_2 f(\pre{b_1'})$.
By definition of $R_2$, 
$\fct \in R_2$ ensures that $p_1' \beta p_2'$ implies $f(\pre{b_1'}) = \pre{b_2'}$; thus, 
 $f(e) \preceq_2'  f'(e_1)$, as required, because $f(e) \preceq_2 f(\pre{b_1'}) = \pre{b_2'} \preceq_2' e_2 = f'(e_1)$. 
 
Summing up, we have that $\oimctp \in R_1$, for $i = 1,2$, $\pi'_i$ is a process of $N(k_{0i})$, $f'$ is an isomorphism 
between $\mathsf{E}_{C_1'}$ and $\mathsf{E}_{C_2'}$, $init(N(k_{01})) \llbracket \sigma_1  e_1 , \delta_{1}' |\rangle \oimp{1}$
is a process sequence for $\pi_1'$, $init(N(k_{02})) \llbracket \sigma_2  e_2, \delta_{2}' |\rangle \oimp{2}$ is a process sequence for $\pi_2'$,
$f'(\sigma_1 e_1) = \sigma_2 e_2$, 
and moreover, for $\beta'$ defined as follows:
    \begin{equation*} 
        \begin{split}
        & \forall p_1 \in k'_1, \forall p_2 \in k'_2: 
          p_{1} \mathrel{\beta}' p_{2} \text{ if and only if } \\
         &\qquad \exists b_1 \in Max(C'_1) \text{ such that }\delta_1'(b_1) = p_1 \text{ and}\\
         &\qquad \exists b_2 \in Max(C'_2) \text{ such that }\delta_2'(b_2) = p_2 \text{ and}\\
         & \qquad \text{either } b_1 \in Min(C'_1) \wedge b_2 \in Min(C'_2) \\
         & \qquad \text{or } \pre{b_1} \neq \emptyset \wedge
         \pre{b_2} \neq \emptyset \wedge 
         f'( \pre{b_1}) = \pre{b_2}
         \},
        \end{split}
    \end{equation*}
we get that $\fctp \in R_2$ by definition of $R_2$.
To complete the proof, we need to check that the definition of $\beta'$ in the triple $\oimctp \in R_1$,
from Definition \ref{oim-bis} of OIM bisimulation,
is coherent with the one obtained from $R_2$ for the triple $(\pi_1', f', \pi_2') \in R_2$, i.e., that:\\

$\forall b_1 \in Max(C_1') \, , \forall b_2 \in Max(C_2'). \;
p_{1} \in k_1' \, , \, p_{2} \in k_2' \, 
\text{ where }
\delta_1'(b_1) = p_1 \text{ and } \delta_2'(b_2) = p_2, $
    \[   
        p_1 \mathrel{\beta'} p_2 \text{ (as by $R_2$)} \iff
         \begin{cases}
            &p_{1} \in \old{1} \, , \, p_{2} \in \old{2} \text{ and }
            p_{1} \mathrel{\beta} p_{2}  \,\qquad \text{(i)}\\
            &\text{or} \\
            &p_{1} \in \generated{1} \, , \, p_{2} \in \generated{2} 
            \qquad\qquad \text{(ii)}
         \end{cases}
    \] 
    
\noindent            
We prove the two implications separately.
        \begin{itemize}
            \item[] {\it Proof $\xRightarrow{}$)} 
            by cases on the definition of $\beta'$:
                \begin{itemize}
                    \item[-] if $\pre{b_1} = \emptyset \wedge 
                        \pre{b_2} = \emptyset$ :\\
                        Then $b_1 \in Min(C_1')$ and 
                        $b_2 \in Min (C_2')$. 
                        For this reason, 
                        $p_{1} \in \old{1}$, 
                         $p_{2} \in \old{2}$ and
                         $p_{1} \mathrel{\beta} p_{2}$,
                        satisfying condition (i).

                      \item[-] if $\pre{b_1} \neq \emptyset \wedge
                                \pre{b_2} \neq \emptyset \wedge 
                                f'(\pre{b_1}) = \pre{b_2}$: \\
                        There are two cases for the event which generates $b_1$:
                        \begin{itemize}
                            \item[+] if $\pre{b_1} = e_1$:
                                then since $f'(\pre{b_1}) = e_2$, 
                                we have $\pre{b_2} = e_2$;
                                hence $p_{1} \in \generated{1}$
                                and $p_{2} \in \generated{2}$, 
                                satisfying condition (ii).
                            
                               \item[+] if $\pre{b_1} \neq e_1$:
                                then, $p_1 \in \old{1}$ and, since $\pre{b_1} \neq \emptyset$,
                                there exists $e_1'$, that occured before $e_1$,
                                such that $\pre{b_1} = e_1'$.
                                By the fact that $f'$ is an isomorphism
                                between $\mathsf{E}_{C_1'}$ and
                                $\mathsf{E}_{C_2'}$, 
                                there exists also $e_2'$, that occurred before $e_2$ (hence, $p_2 \in \old{2}$),
                                where $f'(e_1') = e_2'$ 
                                such that $\pre{b_2} = e_2'$.
                                Note that we also have that $f$ is an isomorphism between 
                                $\mathsf{E}_{C_1}$ and
                                $\mathsf{E}_{C_2}$ such that $f(\pre{b_1}) = \pre{b_2}$.
                                Hence, 
                                we have $p_{1} \mathrel{\beta} p_{2}$,
                                satisfying condition (i).
                        \end{itemize}
                    \end{itemize}
                          
             \item[] {\it Proof $\xLeftarrow{}$)}
                by cases:
                \begin{itemize}
                    \item[-] if $p_{1} \in \old{1}$ and
                        $p_{2} \in \old{2}$ and
                        $p_{1} \mathrel{\beta} p_{2}$: \\
                        then there are two possible cases for $b_1$:
                        \begin{itemize}
                            \item[+] if $b_1 \in Min(C_1')$: \\
                             then, since $p_1$ does not move, $b_1 \in Min(C_1)$ and,
                             since $p_{1} \mathrel{\beta} p_{2}$, also
                             $b_2 \in Min(C_2)$.
                             
                            \item[+] if $b_1 \not\in Min(C_1')$: \\
                             then $\pre{b_1} \neq \emptyset$; however,
                             since $b_1$ does not move, because $p_{1} \in \old{1}$ (and also $p_{2} \in \old{2}$), we have that, 
                             due to $\beta$,  $\pre{b_2} \neq \emptyset$
                             and $f(\pre{b_1}) = \pre{b_2}$,
                             and by conservative extension of $f$,
                             $f'(\pre{b_1}) = \pre{b_2}$.
                        \end{itemize}
                        
                      \item[-] if $p_{1} \in \generated{1} $ and
                        $p_{2} \in \generated{2}$: \\
                        then $\pre{b_1} = e_1$ and $\pre{b_2} = e_2$
                        and $f'(e_1) = e_2$ by definition of $\beta'$, since they are maximal.
                \end{itemize}
        
        \end{itemize}

As mentioned above, the case in which $\pi_2$ moves first is symmetrical and so omitted. 
Therefore, $R_2$ is an fc-bisimulation and $m_{01} \sim_{fc} \; m_{02}$.
\qed
\end{thm}

\begin{thm} {\bf (OIM-bisimilarity and FC-bisimilarity coincide)}\label{oim=fc}
Let $N = (S,A,T)$ be a net and $m_1, m_2$ two markings of $N$.
$m_{1} \; \sim_{oim} \; m_{2}$ if and only if $m_{1} \; \sim_{fc} \; m_{2}$.

\proof
By Theorems \ref{fc-implies-oim-bounded} and \ref{oim-implies-fc-bounded}, we get the thesis.
\qed
\end{thm}

\begin{thm} \label{fc-decidable-bounded}
{\bf (FC-bisimilarity is decidable for finite bounded nets)}
Given $N(m_1)$ and $N(m_2)$ bounded nets, it is decidable to check whether $m_1 \sim_{fc} m_2$.

\proof
By Theorem \ref{oim=fc}, it is enough to check whether there exists an OIM bisimulation $\mathfrak{B}$
for the given net $N$ and initial indexed markings $k_{01}$ and $k_{02}$, with $\alpha(k_{0i}) = m_i$ for $i = 1, 2$.
If we restrict $\mathfrak{B}$ to 
$\mathfrak{B'} = \{\oimct \in \mathfrak{B} \mid 
(k_i, \leq_i) \in \imtrans{init(N(k_{0i}))} \text{ for } i = 1,2\}$
we have that $\mathfrak{B'}$ is still an OIM bisimulation for $m_1, m_2$.
Indeed, by definition $init(N(k_{0i})) \in \imtrans{init(N(k_{0i}))}$; moreover,
if $\oimct \in \mathfrak{B'}$ and 
$\oim{1} \imtrans{t_1} \oimp{1}$, 
then
it is true that $\oim{2} \imtrans{t_2} \oimp{2}$ and 
$\oimp{i}$ is reachable from $init(N(k_{0i}))$ for $i = 1,2$.

Then, to state that $m_1 \sim_{oim} m_2$, it is enough to consider the
ordered indexed markings contained in $\imtrans{init(N(k_{01}))}$ and $\imtrans{init(N(k_{02}))}$ and,
by Proposition \ref{reachable-oim-finite}, these oims are finitely many.
Moreover, given two reachable ordered indexed markings $(k_1, \leq_1)$ and $(k_2, \leq_2)$,
there are finitely many relations $\beta \subseteq k_1 \times k_2$ to consider, as $k_1$ and $k_2$ are finite objects.
Therefore, we can check by exhaustive search whether one of the finitely many possible finite sets 
of triples of type $\oimct$ is an OIM bisimulation.
\qed
\end{thm}

We conclude this section with some comments on the complexity of the decision procedure. 
Assume that the net has $s$ places, $t$ transitions and it is $h$-bounded. 
Then there will be at most $hs$ tokens in every reachable marking, 
and since the possible preorders on $hs$ elements are 
$2^{O (hs \cdot log(hs))}$, 
there are at most 
$2^{O (hs \cdot log(hs))}$
ordered indexed markings.
Since $\beta$ is a binary relation on tokens, it contains at most $O((hs)^2)$ elements; 
therefore, there are at most 
$2^{O (hs \cdot log(hs))}$
possible elements of $\mathfrak{B}$.
Note that, according to Definition \ref{token-game-oim}, it is possible
to construct a labeled transition system
where states are ordered indexed markings
and transitions are derived from $T$.
Therefore,
it is possible to construct an OIM bisimulation
starting from the labeled transistion system containing 
$init(N(k_{01}))$ and $init(N(k_{02}))$.
The algorithm consumes all reachable states of the transition system; 
for each pair of triples, it
requires scanning $O(t^2 (hs)^2)$ transitions for the bisimulation game
(because the transition relation on ordered indexed markings is nondeterministic)
and $O((hs)^3)$ tokens for the condition on $\beta$. 
Therefore the upper bound for our decision procedure is
$2^{O (hs \cdot log(hs) + log(t))}$.
Note that our exhaustion algorithm has no worse time complexity than other proposed algorithms \cite{MP97,JM96}.

\section{I-causal-net Bisimilarity is Decidable} \label{icn-dec-sec}

In the same fashion as in the previous section, we now prove that also i-causal-net bisimilarity is decidable
by defining a new, decidable equivalence based on ordered indexed markings and showing that it coincides
with i-causal-net bisimilarity.

As in the previous section, an OIMC bisimulation is a relation composed of triples of the form $\oimct$,
where the related indexed markings must have the same size (i.e.,  $\size{k_1} = \size{k_2}$)
and the two matching transitions not only must have the same label, but also 
must consume individual tokens related via $\beta$.
However, here we are concerned not only with precedences in individual token generation,
but also in individually matching each consumed token; this means that
we require that the tokens consumed by the first transition are in a {\em bijective
correspondence} via $\beta$ to those consumed by the second one, so that the same causal nets are really generated.
This is a stronger condition than the one of Definition \ref{oim-bis} which,
inspired by Vogler's proof in \cite{Vog91}, only related tokens up-to their generation.

\begin{defi}\label{oimc-bis}{\bf (OIMC bisimulation) }
Let $N = (S, A, T)$ be a P/T net. An OIMC bisimulation is a relation 
$\mathfrak{B} \subseteq OIM(N) \times OIM(N) \times \mathcal{P}( (S \times \nat) \times (S \times \nat))$ 
such that if $\oimct \in \mathfrak{B}$, then:
\begin{itemize}
    \item $\size{k_1} = \size{k_2}$
    
    \item $\forall t_1, k_1', \leq_1'$ if $\oimmvs{1}{t_1}$ 
    (where we assume that $\kboilerplateindex{t_1}{1}$), 
    then there exist $t_2, k_2', \leq_2'$ 
    (where we assume 
    $\kboilerplateindex{t_2}{2}$),
    and for $\beta'$ defined as:
    $\forall p_{1} \in k_1', \forall p_{2} \in k_2'$
    
    \[   
    p_{1} \; \beta' \; p_{2} \iff
         \begin{cases}
            &p_{1} \in k_1'' , \, p_2 \in k_2'' \text{ and } 
            p_{1} \; \beta \; p_{2} \\
            &\text{or}\\
            &p_{1} \in k_1' \setminus k_1'' \, , \, p_{2} \in k_2' \setminus k_2''
         \end{cases}
    \]
    
        the following hold:
        \begin{itemize}
            \item[-] $\oimmvs{2}{t_2}$ where $\oimctp \in \mathfrak{B}$ and $l(t_{1}) = l(t_{2})$ and
            \item[-] $\beta$ contains a bijection from $(k_1 \setminus k_1'')$ to $(k_2 \setminus k_2'')$, i.e.,
            there is a bijection $g: k_1 \setminus k_1'' \rightarrow k_2 \setminus k_2''$ such that if $g(p_1) = p_2$, 
            then $p_1 \beta p_2$.
        \end{itemize}
    \item symmetrically, if $(k_2, \leq_2)$ moves first.
\end{itemize}
Two markings $m_1$ and $m_2$ of $N$ are OIMC bisimilar, 
denoted $m_1 \sim_{oimc} m_2$, 
if there exists an OIMC bisimulation $\mathfrak{B}$ 
containing the triple 
$(init(N(k_{01})), init(N(k_{02})), k_{01} \times k_{02})$ 
where, for $i = 1, 2$, $k_{0i}$ is the initial (i.e., closed) indexed marking such that $m_i = \alpha(k_{0i})$.
\end{defi}

Note that two matching transitions must the same preset size (by the condition on the existence of a bijection $g$ from  
$(k_1 \setminus k_1'')$ to $(k_2 \setminus k_2'')$);
moreover, since $\size{k_1} = \size{k_2}$ and $\size{k_1'} = \size{k_2'}$, we have that the two matching transitions must have the 
same postset size, i.e., $\size{k_1' \setminus k_1''} = \size{k_2' \setminus k_2''}$.

We prove that OIMC-bisimilarity and i-causal-net bisimilarity coincide on P/T nets, by first
showing that i-causal-net bisimilarity implies OIMC-bisimilarity,
and then by showing that OIMC-bisimilarity implies i-causal-net bisimilarity.

\begin{thm}\label{cn-implies-oimc} {\bf (ICN-bisimilarity implies OIMC-bisimilarity)}
Let $N = (S,A,T)$ be a net. Given two markings $m_{01}, m_{02}$ of $N$,
if $m_{01} \sim_{icn} m_{02}$, then $m_{01} \sim_{oimc} m_{02}$. 
\proof

If $m_{01} \; \sim_{icn} \; m_{02}$, then there exists an icn-bisimulation $R_1$ 
containing a triple $(\rho^0, \,C^0, \, \rho^0)$, 
where $C^0$ contains no events, $\rho^0 = \emptyset$ is undefined for all $b \in C^0$ 
and, for $i = 1, 2$, $(C^0, \rho^0)$ is a partial process of $N(m_{0i})$ for $m_{0i}$ (i.e., this is the same as requiring that
$|Max(C^0)| = |m_{01}| = | m_{02}|$).
Given $k_{0i}$ closed indexed marking such that $m_{0i} = \alpha(k_{0i})$ for $i = 1, 2$, let us consider 
    \begin{equation*} \label{R2-OIMC}
        \begin{split}
        R_2 \overset{def}{=} \lbrace \oimct | & \cnt \in R_{1} \text{ and, for } i = 1,2, \\
         & \pi_i = (C, \rho_i) \text{ is a partial process of  $N(k_{0i})$} \text{ and} \\ 
          & init(N(k_{0i})) \llbracket \sigma, \delta_i |\rangle \oim{i} \text{ is a process sequence for } \pi_i \text{ and}\\
         & \forall p_1  \in k_1 , \text{ with $b_1.\delta_1(b_1) = p_1,$} \; \forall p_2 \in k_2 , \text{ with $b_2.\delta_2(b_2) = p_2$,}\\
         &  \text{we have that: }\;  p_1 \beta p_2 \text{ if and only if } 
         \pre{b_1} = \pre{b_2} 
         \rbrace .
        \end{split}
    \end{equation*}
If we prove that $R_2$ is an OIMC bisimulation, then, as $(\rho^0, \,C^0, \, \rho^0) \in R_1$ and, for $i = 1, 2$, we have that
$\pi_{0i} = (C^0, \rho_0)$ is a partial process for $N(k_{0i})$, 
$init(N(k_{0i})) \llbracket \epsilon, \delta_{0i} |\rangle init(N(k_{0i}))$ 
is a process sequence for $\pi_{0i}$
and $\forall b_1, b_2 \in C^0$ we have $\pre{b_1} = \emptyset = \pre{b_2}$, it follows that 
$(init(N(k_{01})), init(N(k_{02})), k_{01} \times k_{02}) \in R_2$ by definition of $R_2$
and, therefore, $m_{01} \mathrel{\sim_{oimc}}  m_{02}$.    

Assume $\oimct \in R_2$. By symmetry, we consider only the case when $(k_1, \leq_1)$ moves first.
Let $\oimmvs{1}{t_1}$.

By definition of $R_2$, a process sequence
$init(N(k_{01})) \llbracket \sigma, \delta_1 |\rangle \oim{1}$ for $\pi_1 = (C, \rho_1)$ exists.
By Theorem \ref{c-leq-proc-oim-3-4} it follows that $\pi_1 = \proctransp{1}{e} = \pi_1'$ 
where $\rho'_1(e) = t_1$ and, moreover, that
$init(N(k_{01})) \llbracket \sigma e, \delta_1' |\rangle \oimp{1}$ is a process sequence for $\pi_1'$.
Since $\cnt \in R_1$, it follows that
$\pi_2 = \proctransp{2}{e} = \pi_2'$,
where $\rho'_2(e) = t_2$ and
$\cntp \in R_1$.
By definition of $R_2$, a process sequence
$init(N(k_{02})) \llbracket \sigma, \delta_2 |\rangle \oim{2}$ for $\pi_2$ exists; moreover, 
$init(N(k_{02})) \llbracket \sigma e, \delta_2' |\rangle \oimp{2}$ is a process sequence for $\pi_2'$
(as $\pi_2$ and $\pi_2'$ are partial processes of $N(k_{02})$). Hence,
by Theorem \ref{c-leq-proc-oim-3-4} we have $\oimmvs{2}{t_2}$.

Summing up, we have that $\cntp \in R_1$, for $i = 1, 2$, $\pi_i'$ is a partial process of $N(k_{0i})$, 
$init(N(k_{01})) \llbracket \sigma e, \delta_1' |\rangle \oimp{1}$ is a 
process sequence for $\pi_1'$,
$init(N(k_{02})) \llbracket \sigma e, \delta_2' |\rangle \oimp{2}$ is a process sequence for $\pi_2'$, 
and moreover, for $\beta'$ defined as follows:
    \begin{equation*}
        \begin{split}
        & \forall p_1 \in k'_1, \text{ with $b_1.\delta_1'(b_1) = p_1,$}\ \; \forall p_2 \in k_2' , \text{ with $b_2.\delta_2'(b_2) = p_2$,}\\
         &  \text{we have that: }\;  p_1 \beta' p_2 \text{ if and only if } 
         \pre{b_1} = \pre{b_2},
        \end{split}
    \end{equation*}
we get that $\oimctp \in R_2$ by definition of $R_2$.

Therefore, we have proved that to the move $\oimmvs{1}{t_1}$, $(k_2, \leq_2)$ can reply with the move
$\oimmvs{2}{t_2}$, where $\oimtp \in R_2$ and $l(t_1) = l(t_2)$.
Hence, in order to prove that $\oimt$ is a OIMC bisimulation triple, as required, 
it remains to prove that 
the definition of $\beta'$ arising from $R_2$, i.e., the unique $\beta'$ such that the triple $((k_1', \leq_1'), (k_2', \leq_2'), \beta') \in R_2$
for $(\rho'_1, C', \rho_2') \in R_1$, 
is coherent with the one of Definition \ref{oimc-bis}, 
i.e., it implies both
    \begin{enumerate}
        \item $\forall p_{1} \in k_1' \, , \, \forall p_{2} \in k_2' \, $
    \[   
        p_{1} \mathrel{\beta'} p_{2} \text{ (as by $R_2$)}\iff
         \begin{cases}
            &p_{1} \in k_1''\, , \, p_{2} \in k_2'' \text{ and }
            p_{1} \; \beta \; p_{2} \quad\qquad \text{(i)}\\
            &\text{or} \\
            &p_{1} \in k_1' \setminus k_1'' \, , \,
            p_{2} \in k_2' \setminus k_2'' 
            \qquad\qquad\quad \text{(ii)}
         \end{cases}
    \]
 and
        \item there is a bijection $g: k_1 \setminus k_1'' \rightarrow k_2 \setminus k_2''$ such that if $g(p_1) = p_2$, 
            then $p_1 \beta p_2$.
    \end{enumerate}
                   
\begin{itemize}
    \item[] {\it Proof 1)} \\
        The two implications are proved separately.
        \begin{itemize}
            \item if $\delta_1'(b_1) = p_{1} \mathrel{\beta'} p_{2} = \delta_2'(b_2)\; \iff \; 
           \pre{b_1} \, = \, \pre{b_2}$: \\
            There are four possibilities for $p_{1}, p_{2}$:
            \begin{itemize}
                \item[-] if $p_{1} \in \generated{1}$ and $p_{2} \in \generated{2}$: condition (ii) is trivial.
                
                \item[-] if $p_{1} \in \old{1}$ and $p_{2} \in \old{2}$: then $\delta_1'(b_1) = \delta_1(b_1) = p_{1}$ and $\delta_2'(b_2) = 
                \delta_2(b_2) = p_{2}$,
                so that,
                since $\oimct \in R_2$, the hypothesis $\pre{b_1} \, = \, \pre{b_2}$ ensures that $p_1 \mathrel{\beta} p_2$ holds. 
                Then, condition (i) is satisfied.
                
                \item[-] other cases: absurd, because $\pre{b_1} \, = \, \pre{b_2}$.
            \end{itemize}
 
              \item if (i) or (ii) hold: 
                \begin{itemize}
                  \item[-] if (i) holds: since $\oimct \in R_2$, we have that $p_1 \beta p_2$ iff $\pre{b_1} = \pre{b_2}$
                      for $p_i = \delta_i(b_i)$ for $i = 1, 2$. Since $p_{1}, p_{2}$ do not move, we have that
                      $p_i = \delta_i'(b_i)$ for $i = 1, 2$, and then, since $\pre{b_1} = \pre{b_2}$,
                      the thesis
                       $p_{1} \mathrel{\beta'} p_{2}$ follows.
                    
                    \item[-] if (ii) holds: 
                       Since $p_1$ and $p_2$ are generated, we have that there exists an event $e$ and two conditions $b_1$ and $b_2$
                       such that $e = \pre{b_1} = \pre{b_2}$ with 
                       $\delta_i'(b_i) = p_i$ for $i = 1, 2$.
                        Hence, $p_{1} \mathrel{\beta'} p_{2}$.
                \end{itemize}
        \end{itemize}
        
      \item[] {\it Proof 2)} \\
        Note that, for $i = 1, 2$, $\delta_i$ maps the preset of $e$ bijectively to the tokens in the preset of $t_i$.
        Hence, if $g$ maps each $\delta_1(b)$ to $\delta_2(b)$ for each $b \in \pre{e}$, then this is a bijection
        from $k_1 \setminus k_1''$ to $k_2 \setminus k_2''$. Since $\pre{b} = \pre{b}$, then 
        we have that $\delta_1(b)$ $\beta$ $\delta_2(b)$ by the choice of $\oimct \in R_2$.
      
         \end{itemize}

Note that $\size{k_1'} = \size{k_2'}$, as we already have 
$\size{k_1} = \size{k_2}$, $\size{\deleted{1}} = \size{\deleted{2}}$
 and  $\size{\generated{1}} =$ $ \size{\post{e}} = \size{\generated{2}}$.
Therefore, not only we have proved that $\oimctp \in R_2$, but also that the triple $\oimct$ is an OIMC bisimulation triple, as required.

As mentioned above, the case in which $(k_2, \leq_2)$ moves first is symmetrical and so omitted.
Therefore, $R_2$ is an OIMC bisimulation, and thus $m_{01} \; \sim_{oimc} \; m_{02}$.
\qed
\end{thm}    

\begin{thm} \label{oimc-implies-cn} {\bf (OIMC-bisimilarity implies ICN-bisimilarity)}
Let $N = (S,A,T)$ be a net.
Given two markings $m_{01}, m_{02}$ of $N$,
if $m_{01} \; \sim_{oimc} \; m_{02}$, then $m_{01} \; \sim_{icn} \; m_{02}$.

\proof
If $m_{01} \; \sim_{oimc} \; m_{02}$, then there exists an OIMC bisimulation $R_1$ containing the tuple 
$(init(N(k_{01})),  init(N(k_{02})), k_{01} \times k_{02})$, 
where $\alpha(k_{01}) = m_{01}$, $\alpha(k_{02}) = m_{02}$, and $k_{01}, k_{02}$ are closed.
 Let us consider 
    \begin{equation*} \label{R2-I-CN}
        \begin{split}
       R_2 \overset{def}{=} \lbrace \cnt | & \oimct \in R_{1} \text{ and, for } i = 1,2, \\
          & \pi_i = (C, \rho_i) \text{ is a partial process of } N(k_{0i}) \text{ and} \\ 
          & init(N(k_{0i})) \llbracket \sigma, \delta_i |\rangle \oim{i} \text{ is a process sequence for } \pi_i \text{ and}\\
          & \forall p_1  \in k_1 , \text{ with $b_1.\delta_1(b_1) = p_1,$} \; \forall p_2 \in k_2 , \text{ with $b_2.\delta_2(b_2) = p_2$,}\\
          &  \text{we have that: } \; p_1 \beta p_2 \text{ if and only if } \pre{b_1} = \pre{b_2}
                \rbrace .
        \end{split}
    \end{equation*}
    
Note that $(\rho^0, C^0, \rho^0) \in R_2$, 
where $C^0$ contains no transitions and $\rho^0$ is undefined for all $b \in C^0$,
because 
$(init(N(k_{01})), init(N(k_{02})), k_{01} \times k_{02}) \, \in R_1$ and, for $i = 1, 2$,
$\pi_{0i} = (C^0, \rho^0)$ is a partial process of $N(m_{0i})$ and
$init(N(k_{0i})) \llbracket \epsilon, \delta_{0i} |\rangle init(N(k_{0i}))$ 
is a process sequence for $\pi_{0i}$. Therefore, if we prove that $R_2$ is an icn-bisimulation, 
since $(\rho^0, C^0, \rho^0) \in R_2$, we have $m_{01} \mathrel{\sim_{icn}} m_{02}$.

Assume $\cnt \in R_2$, where $\pi_1 = (C, \rho_1)$
and $\pi_2 = (C, \rho_2)$.  By symmetry, we consider only the case when $\pi_1$ moves first.
Let $\pi_1 = \proctransp{1}{e} = \pi_1'$, where $\rho'_1(e) = t_1$.

By definition of $R_2$, there exists a process sequence 
$init(N(k_{01})) \llbracket \sigma, \delta_1 |\rangle \oim{1}$ for $\pi_1$; moreover,
$init(N(k_{01})) \llbracket \sigma e, \delta_1' |\rangle \oimp{1}$ is a process sequence for $\pi_1'$.
Hence, by Theorem \ref{c-leq-proc-oim-3-4} we have that $\oimmvs{1}{t_1}$.

Since $\oimct \in R_1$, it follows that  there exist $t_2, k_2', \leq_2', \beta'$ such that
$\oimmvs{2}{t_2}$, where $((k_1', \leq_1'),$ $(k_2', \leq_2'), $ $ \beta') \in R_1$.
By definition of $R_2$, there exists a suitable process sequence 
$init(N(k_{02})) \llbracket \sigma, \delta_2 |\rangle \oim{2}$ for $\pi_2$. Hence, by Theorem \ref{c-leq-proc-oim-3-4}, it follows
that there exists some event $\overline{e}$ such that 
$\pi_2 = (C, \rho_2) \derivp{\overline{e}} (\overline{C}', \rho_2') = \pi_2'$, where $\rho'_2(\overline{e}) = t_2$ and that
$init(N(k_{02})) \llbracket \sigma \overline{e}, \delta_2' |\rangle \oimp{2}$ is a process sequence for $\pi_2'$.

We want to argue that event $\overline{e}$ can be chosen to be exactly event $e$, so that $\overline{C}' = C'$.
In fact, as $\oimct \in R_1$, we know that there exists a bijection $g$ from the tokens consumed by $t_1$ to 
the tokens consumed by $t_2$ such that  
if $g(p_1) = p_2$, then $p_1\beta p_2$. By definition of $R_2$, we have that 
if $\delta_1(b_1) = p_1$ and $\delta_2(b_2) = p_2$,
          then $p_1 \beta p_2$ if and only if $\pre{b_1} = \pre{b_2}$. Since the definition of $\delta_2$ 
          on conditions
          generated by the same event is arbitrary (i.e., any choice is fine), we can partially redefine it by taking $\delta_2(b_1) = p_2$
          so that token $p_1 \in  k_1 \setminus k_1''$ and token $p_2 \in k_2 \setminus k_2''$, such that $g(p_1) = p_2$, 
          are the image, via $\delta_1$ and 
          $\delta_2$ respectively,  of the same condition $b_1$.
          Iterating this procedure for all the pairs of tokens in the bijection $g$, we get that $\pre{e} = \pre{\overline{e}}$.
          Moreover, the label of $e$ and $\overline{e}$ is the same, as $l(t_1)= l(t_2)$. Finally, since the sets of the generated tokens 
          $k_1' \setminus k_1''$
          and $ k_2' \setminus k_2''$ have the same size and the choice of the postset
          of $\overline{e}$ is completely arbitrary, we can take $\post{\overline{e}} = \post{e}$.

Summing up, we have proved that $\oimctp \in R_1$, for $i = 1, 2$, $\pi'_i$ is a partial process of $N(k_{0i})$, 
$init(N(k_{01})) \llbracket \sigma  e , \delta_{1}' |\rangle \oimp{1}$
is a process sequence for $\pi_1'$, $init(N(k_{02})) \llbracket \sigma  e, \delta_{2}' |\rangle \oimp{2}$ is a process 
sequence for $\pi_2'$,
and, moreover, for $\beta'$ defined as follows:
    \begin{equation*}
        \begin{split}
        & \forall p_1 \in k'_1, \text{ with $b_1.\delta_1'(b_1) = p_1,$}\ \; \forall p_2 \in k_2' , \text{ with $b_2.\delta_2'(b_2) = p_2$,}\\
         &  \text{we have that: }\;  p_1 \beta' p_2 \text{ if and only if } 
         \pre{b_1} = \pre{b_2},
        \end{split}
    \end{equation*}
we get that $\cntp \in R_2$ by definition of $R_2$.
To complete the proof, we need to check that the definition of $\beta'$ in the triple $\oimctp \in R_1$,
from Definition \ref{oimc-bis} of OIMC bisimulation,
is coherent with the one obtained from $R_2$ for the triple $\cntp \in R_2$, i.e., that the following condition holds:\\

$\forall b_1 \in Max(C'), \forall b_2 \in Max(C') \, . \,
p_{1} \in k_1' \, , \, p_{2} \in k_2' \, 
\text{ with }
\delta_1'(b_1) = p_1, 
\delta_2'(b_2) = p_2 $
    \[   
        p_1 \mathrel{\beta'} p_2 \text{ (as by $R_2$)} \iff
         \begin{cases}
            &p_{1} \in \old{1} \, , \, p_{2} \in \old{2} \text{ and }
            p_{1} \mathrel{\beta} p_{2}  \,\qquad \text{(i)}\\
            &\text{or} \\
            &p_{1} \in \generated{1} \, , \, p_{2} \in \generated{2} 
            \qquad\qquad \text{(ii)}
         \end{cases}
    \] 

\noindent
We prove the two implications separately.

       \begin{itemize}
        \item[] {\it Proof $\xLeftarrow{}$)} 
        Assume $p_1 = \delta_1'(b_1)$ and $p_2 = \delta_2'(b_2)$.  
        We prove the thesis by assuming (i) or (ii) above:
            \begin{itemize}
  
                  \item[-] if $p_{1} \in \old{1}$ and $p_{2} \in \old{2}$ 
                         and $p_{1} \; \beta p_{2}$: \\
                         Since the tokens do not move, and also  $\oimct \in R_1$, 
                         we have $p_{1} \; \beta \; p_{2} \; \Leftrightarrow{} \; $
                         $\pre{b_1} =\pre{b_2}$.  
                 \item[-] if $p_{1} \in \generated{1}$ and $p_{2} \in \generated{2}$: \\
                        Since $\proctransp{i}{e}$ where $\rho_i'(e) = t_i$ for $i = 1,2$, 
                        then $t_1$ and $t_2$ are mapped on the same event $e$.
                        Therefore $\pre{b_1} \; = \; e \; =  \pre{b_2}$.

            \end{itemize}
                    
         \item[] {\it Proof $\xRightarrow{}$)} Consider the event $e$, that is an event of $C'$ and not of $C$, 
                such that $\rho_1'(e) = t_1$ and $\rho_2'(e) = t_2$.
              There are four possibilities for $p_{1} = \delta_1'(b_1)$ and $p_{2} = \delta_2'(b_2)$ such that $\pre{b_1} = \pre{b_2}$:
             \begin{itemize}
                \item[-] if $p_{1} \in \old{1}$ and $p_{2}\in \old{2}$:\\
                         Since the tokens did not move, and $\oimct \in R_1$, 
                         we have $\pre{b_1} = \pre{b_2} \; \Leftrightarrow{} \;$ 
                         $p_{1} \mathrel{\beta} p_{2} $. Therefore, condition (i) holds.
                         
              \item[-] if $p_{1}\in \generated{1}$ and $p_{2}\in \generated{2}$: \\
                      then $p_{1} \mathrel{\beta'} p_{2}$ by condition (ii).
                         
                \item[-] other cases: 
                    absurd, since $\pre{b_1} = \pre{b_2}$. 
            \end{itemize}
    \end{itemize}    


As mentioned above, the case in which $\pi_2 = (C, \rho_2)$ moves first is symmetrical and so omitted.
Therefore, $R_2$ is an icn-bisimulation and so $m_{01} \mathrel{\sim_{icn}} \; m_{02}$.
\qed
\end{thm}
     
 \begin{thm} {\bf (OIMC-bisimilarity and ICN-bisimilarity coincide)}\label{oimc=cn}
Let $N = (S,A,T)$ be a net and $m_1, m_2$ two markings of $N$.
$m_{1} \; \sim_{oimc} \; m_{2}$ if and only if $m_{1} \; \sim_{icn} \; m_{2}$.
\proof
By Theorems \ref{cn-implies-oimc} and \ref{oimc-implies-cn}, we get the thesis.
\qed
\end{thm}

\begin{thm} \label{cn-decidable-bounded}
{\bf (ICN-bisimilarity is decidable for finite bounded nets)}
Given $N(m_1)$ and $N(m_2)$ bounded nets, it is decidable to check whether $m_1 \sim_{icn} m_2$.
\proof
By Theorem \ref{oimc=cn}, it is enough to check whether there exists an OIMC bisimulation $\mathfrak{B}$
for the given net $N$ and initial markings $m_{1}, m_{2}$ (with indexed initial markings $k_{01}$ and $k_{02}$).
The proof then follows the same steps of Theorem \ref{fc-decidable-bounded}.
\qed
\end{thm}
Note that the complexity of this procedure, being very similar to that discussed at the end of Section \ref{fc-dec-sec}, 
is again $2^{O (hs \cdot log(hs) + log(t))}$.

\begin{thm} \label{icn>fc-th}
{\bf (ICN-bisimilarity is finer than FC-bisimilarity)}
Let $N = (S,A,T)$ be a net and $m_1, m_2$ two markings of $N$.
If $m_{1} \; \sim_{icn} \; m_{2}$, then $m_{1} \; \sim_{fc} \; m_{2}$.
\proof
By Theorem \ref{oimc=cn}, we have that $\sim_{icn}$ coincides with $\sim_{oimc}$.
Note that an OIMC bisimulation is actually also an OIM bisimulation, so that $\sim_{oimc} \subseteq \sim_{oim}$.
By Theorem \ref{oim=fc}, we have that $\sim_{oim}$ coincides with $\sim_{fc}$, so that the thesis
$\sim_{icn} \subseteq \sim_{fc}$ follows trivially. (Example \ref{ex-icn-vs-sfc} shows that the implication is strict.)
\qed
\end{thm}

%
\section{Conclusion and Future Research}\label{conc-sec}
%

We have extended Vogler's proof technique in \cite{Vog91}, based on ordered markings, that he used to prove decidability of (strong) fully-concurrent bisimilarity for safe nets, to bounded nets by means of indexed ordered markings. The extension is flexible enough
to be applicable also to other similar equivalences, such as i-causal-net bisimilarity, a novel behavioral equivalence slightly coarser than
causal-net bisimilarity \cite{vG15,Gor22}.
While decidability of fully-concurrent bisimilarity for bounded nets was already proved
by Montanari and Pistore \cite{MP97}, our result for i-causal-net bisimilarity is, of course, new.

However, the approach of \cite{MP97} is not defined directly on Petri nets, rather
it exploits an encoding of Petri nets into so-called {\em causal automata}, a model of computation designed for handling dependencies between transitions by means of names.  
In addition to this, their encoding works modulo isomorphisms, so that, in order to handle correctly the dependency names, at each step of the construction costly renormalizations are required.
Along the same line, recently {\em history-dependent automata} \cite{BMS15a,BMS15b} have been proposed. 
They are a much refined version of causal automata, 
retaining not only events but also their causal relations.
Moreover, they are equipped with interesting categorical properties 
such as having symmetry groups over them, which allow for state reductions. 
As in the former work, the latter ones do not work directly on the net and may require minimizations
(albeit \textit{automatic}, in the case of HD automata).
On the contrary, our construction is very concrete and works directly on the net. Thus, we conjecture that, even if the worst-case complexity is roughly the same, our algorithm may perform generally better.

Decidability of fully-concurrent bisimilarity using the ordered indexed marking idea
was claimed to have been proved by Valero-Ruiz in his PhD thesis \cite{VR93} for the subclass of bounded P/T nets where transitions pre- and post-sets are sets. Valero-Ruiz's approach differs from ours both in how the proof is conducted and in accuracy.
In his work, ordered indexed markings are defined in such a way that they are always closed, but
depending on the chosen token to remove, there may appear a hole in the indexing (cf. Example \ref{token-game-oim-example}), and therefore it is stated that the resulting ordered indexed marking may be subject to renaming to be again closed.
This definition does not ensure the individuality of tokens: one token not used in a transition can be renamed, so that (even if it is not taking part to the transition) its index before and after the transition is different.
Moreover, isomorphism of ordered indexed marking is defined only on closed ones, therefore it is not clear how the renaming is carried on. At the same time, it is left implicit how relation $\leq$ should behave w.r.t. renaming: since the individuality of tokens cannot be assumed, this is not a trivial detail.
Another critical point is in the definition of the indexed ordered marking-based bisimulation (similar to Definition \ref{oim-bis}), where the possible renaming of tokens between transition steps is not taken into account.
These inaccuracies undermine Valero-Ruiz's result on decidability of fully-concurrent bisimilarity for the subclass of bounded P/T nets where transitions pre- and post- sets are sets. Therefore, our work can be considered the first one to have proved it using the ordered indexed marking approach, and on the larger class of bounded nets.

A natural question is whether it is possible to decide these equivalences for larger classes of nets, notably unbounded P/T nets.
However, as Esparza observed in \cite{Esp98}, all the behavioral equivalences ranging from interleaving bisimilarity to
fully-concurrent bisimilarity are undecidable on unbounded P/T nets. So, there is no hope to extend our result about fc-bisimilarity further. Nonetheless, the proof of undecidability by Jan\u{c}ar \cite{Jan95} does not apply to (i-)causal-net bisimilarity, so that
the decidability of (i-)causal-net bisimilarity over unbounded P/T nets is open.

We conclude by offering a panorama of decidability results over the spectrum of
the behavioral equivalences fully respecting causality and the branching time, defined over finite Petri nets, summarized as follows:
\[
\sim_p \; \subseteq \; \sim_{cn} = \sim_{sp} \; \subseteq \; \sim_{icn} = \sim_{oimc} \; \subseteq \; \sim_{sfc} \; \subseteq \; \sim_{fc} = \sim_{oim}
\]
where place bisimilarity $\sim_p$ \cite{ABS91,Gor21} is the finest one, then causal-net bisimilarity $\sim_{cn}$ \cite{vG15,Gor22}
(which is equivalent to structure-preserving bisimilarity $\sim_{sp}$ \cite{vG15}), then i-causal-net bisimilarity
(which is equivalent to OIMC bisimilarity $\sim_{oimc}$), then state-sensitive fully-concurrent bisimilarity 
$\sim_{sfc}$ \cite{Gor22}, and finally
 fully-concurrent bisimilarity $\sim_{fc}$ \cite{BDKP91} (which is equivalent to OIM bisimilarity $\sim_{oim}$), which is the coarsest one.

Place bisimilarity $\sim_p$, originally defined in \cite{ABS91}, is a behavioral equivalence that, differently from all the other listed above,
is based on relations on the finite set of net places, rather than on relations on the (possibly infinite) set of reachable markings. 
This behavioral relation was recently proved decidable for finite unbounded 
P/T nets in \cite{Gor21}. In that paper also a novel variant behavioral equivalence, called d-place bisimilarity $\sim_d$, was 
introduced; this equivalence, which is coarser than $\sim_p$, finer than fully-concurrent bisimilarity $\sim_{fc}$, but incomparable with the other equivalences listed above, is the coarsest decidable behavioral equivalence, fully respecting causality and the branching time, defined so far
for finite unbounded P/T nets. 

Causal-net bisimilarity $\sim_{cn}$ \cite{vG15,Gor22} is decidable on finite bounded Petri nets. This can be proved by exploiting
its equivalent characterization in terms of structure-preserving bisimilarity $\sim_{sp}$ \cite{vG15}. In fact, a structure-preserving
bisimulation may be seen as a relation composed of triples of type (marking, bijection, marking), where the first component and the third component
are reachable markings (which are finitely many for finite bounded nets) and the second component is a bijection between the two (hence, 
this is one in a set of
at most $k!$ bijections, if the size of the markings is $k$). Therefore, as the possible triples can be finitely many for finite bounded P/T nets,
there can only be finitely many candidate relations (which are all finite) to be structure-preserving bisimulations.
As mentioned above, decidability of $\sim_{cn}$ over finite unbounded P/T nets is an open problem.
As discussed in \cite{vG15}, causal-net bisimilarity is the coarsest semantics
respecting {\em inevitability} \cite{MOP89}, i.e., if two
systems are equivalent, and in one the occurrence of a certain action is inevitable, then so is it in the other one.

We have proved that $\sim_{icn}$ is decidable on finite bounded P/T nets by means of the equivalent characterization
in terms of $\sim_{oimc}$. As mentioned above, decidability of $\sim_{icn}$ over finite unbounded nets is an open problem.
I-causal-net bisimilarity is the coarsest semantics respecting the structure of the net, i.e., the coarsest bisimulation-based one ensuring that 
related markings generate the same causal nets.

{\em State-sensitive} fully-concurrent bisimilarity $\sim_{sfc}$ \cite{Gor22} is a slight refinement of fully-concurrent 
bisimilarity requiring that the current markings have the same size. Example \ref{ex-icn-vs-sfc} shows that $\sim_{sfc}$ is coarser than
$\sim_{icn}$.
Hence, even if $\sim_{sfc}$ is the coarsest equivalence to be resource-aware, we think that $\sim_{icn}$ is more accurate,
as an observer that can really observe the distributed state should be able to observe the structure of the transitions.
It is easy to observe that $\sim_{sfc}$ can be decided over finite bounded Petri nets, by simply enhancing the
definition of OIM bisimulation: it is enough to add the condition that the current indexed markings $k_1$ and $k_2$
have the same size, in order to obtain a slightly stronger bisimulation relation, say {\em OIMS bisimulation}, whose induced
behavioral equivalence, say $\sim_{oims}$, is, of course, decidable as well.
Also the decidability of $\sim_{sfc}$ over finite unbounded nets is open,
even if we conjecture that it is undecidable.

Finally, we have proved that fully-concurrent bisimilarity $\sim_{fc}$, which is the coarsest equivalence 
fully respecting causality and the branching time, is decidable for finite bounded nets by means of its characterization in terms of
$\sim_{oim}$, while,
as mentioned above, it is undecidable for finite P/T nets with at least two unbounded places \cite{Jan95,Esp98}.


On BPP nets, i.e., nets whose transitions have singleton preset but whose set of reachable markings can be infinite, the classification 
above is largely simplified, as it is possible to prove  \cite{Gor22} that
\[
\sim_p \; = \; \sim_{cn} \; = \; \sim_{icn} \; = \; \sim_{sfc} \; = \; \sim_t \; \; \subseteq \; \; \sim_{fc} \; = \; \sim_d \; = \; \sim_{ht}
\]
where $\sim_t$ is {\em team bisimilarity} \cite{Gor22}, $\sim_d$ is d-place bisimilarity \cite{Gor21} and $\sim_{ht}$ is {\em h-team bisimilarity} \cite{Gor22}. All these equivalences can be decided for BPP nets in polynomial time.

As a future work, we plan to extend Vogler's results in \cite{Vog95} about decidability of weak fully-concurrent bisimilarity on 
safe nets with silent moves, to bounded nets with silent moves, by means of our indexed marking idea.\\

\noindent
{\bf Acknowledgements:}
We would like to thank the anonymous reviewers for useful comments and suggestions that helped us to improve the presentation of the paper. 
The second author wishes to thank 
Rob van Glabbeek for fruitful discussions on the difference between causal-net bisimilarity and i-causal-net bisimilarity.

\bibliographystyle{alphaurl}
\bibliography{references.bib}

\end{document}